\documentclass[times,twocolumn,final,authoryear]{elsarticle}

\usepackage{jasr}
\usepackage{framed,multirow}

\usepackage{amssymb}
\usepackage{amsmath}
\usepackage{latexsym}
\usepackage{graphicx}
\usepackage{epstopdf}
\usepackage{color}

\usepackage{amssymb}

\newcommand{\gsim}{\gtrsim\!}

\usepackage{url}
\usepackage{xcolor}
\definecolor{newcolor}{rgb}{.8,.349,.1}

\usepackage[citebordercolor=white]{hyperref}

\journal{Advances in Space Research}

\begin{document}

\verso{A.M. Bykov \textit{etal}}

\begin{frontmatter}
\title{Modeling of GeV-TeV gamma-ray emission of Cygnus Cocoon}%
\author[1]{A.M. {Bykov}}
\ead{byk@astro.ioffe.ru}
\author[1]{M.E. {Kalyashova}\corref{cor1}}
\cortext[cor1]{Corresponding author}
  \ead{m.kalyashova@gmail.com}
\address[1]{Ioffe Institute, Saint Petersburg, Polytechnicheskaya str., 26, 194021, Russia}
\received{}
\finalform{}
\accepted{}
\availableonline{© 2022. This manuscript version is made available under the CC-BY-NC-ND 4.0 license \url{https://creativecommons.org/licenses/by-nc-nd/4.0/}}
\communicated{}
\begin{abstract}
OB-associations and superbubbles being energetically essential galactic powerhouses are likely to be the important acceleration sites of galactic cosmic rays (CRs). The emission profile of $\gamma$-ray sources related to superbubbles and stellar clusters indicates on  continuous particle acceleration by winds of massive stars. One of the most luminous galactic $\gamma$-ray sources is Cygnus Cocoon superbubble, observed by multiple instruments, such as {\it Fermi}-LAT, ARGO, and, recently, HAWC. We discuss a model of particle acceleration and transport in a superbubble to explain GeV-TeV $\gamma$-ray spectrum of Cygnus Cocoon, which has a break at the energy of about 1 TeV. It is shown that the $\gamma$ rays produced by hadronic interactions of high-energy protons accelerated by an ensemble of shocks from winds of massive stars and supernovae in the Cygnus Cocoon can explain the observations. 
The proton spectral shape at the highest energies depends on the MHD-fluctuation spectrum in the Cocoon. The viable solutions for Cygnus Cocoon may be applied to some other associations showing similar behaviour. We briefly discuss the similarity and differences of particle acceleration processes in extended superbubbles and compact clusters of young massive stars as represented by Westerlund 1 and 2 $\gamma$-ray sources.  
\end{abstract}
\end{frontmatter}


\section{Introduction}
The galactic star-forming regions are considered to be the likely sources of high-energy cosmic rays up to PeV, as suggested by \citet[][]{1983SSRv...36..173C},    \citet[][]{Bykov2014}, \citet[][]{Lingenfelter18}, \citet[][]{2019ApJ...879...66T}, \citet[][]{2020SSRv..216...42B},  
\citet[][]{2021MNRAS.504.6096M}. 
 Studies of the CR composition show that the volatile elements in CR material are likely accelerated from a plasma of temperature above  $\sim 2$ MK, which is typical for superbubbles produced by massive star winds and supernova explosions \citep[see][for a recent discussion]{2021arXiv210615581T}. The observed overabundance of $^{22}$Ne in galactic CRs can be understood with the models of CR acceleration in the powerful winds of Wolf-Rayet type stars in young massive star clusters \citep[][]{2008NewAR..52..427B,2019JPhCS1400b2011K,2020MNRAS.493.3159G}.

Whether the massive stars associations and clusters can indeed efficiently accelerate CRs may be verified by modeling the spatial and spectral features of $\gamma$-ray emission from galactic OB-associations and compact clusters.  
\citet{AharonianNat2019} using {\it Fermi}-LAT and H.E.S.S. data for OB-association Cygnus OB2 and compact clusters Westerlund 1 and Westerlund 2, examined the spatial distribution of $\gamma$-ray emission from these objects and found that the CR density behaves like $r^{-1}$. They concluded that relativistic particles are continuosly injected into the interstellar medium, which may be an argument for combined action of multiple stellar winds. This supports the idea that massive stellar clusters and OB-associations are likely galactic CR accelerators. Both superbubbles and massive star clusters are likely dominating sources of very-high-energy $\gamma$ rays from the starburst galaxies \citep[see e.g.][]{Ohm16}.

\citet{AckermannSB2011} presented {\it Fermi}-LAT GeV observations of the extended $\gamma$-ray source, associated with the superbubble surrounding the Cygnus OB2 region, 
known as Cygnus Cocoon. 
They pointed out that the Cocoon must be the site of the active particle acceleration. Cygnus Cocoon is located at $\sim 1.4$ kpc from Solar system, with a size of $\sim 55$ pc and a total mechanical power exceeding $10^{38}$ erg/s  \citep[][]{AharonianNat2019, AckermannSB2011, Hanson2003}. According to {\it Fermi}-LAT analysis, at GeV energies spectral energy distribution has a power-law form with a spectral index of -2.1. Studies of this region at TeV energies  started with  HEGRA  \citep{2002A&A...393L..37A} and continued with Milagro \citep[][]{Abdo2007a,Abdo2007b,Abdo2012}, ARGO \citep[]{Bartoli2014}, VERITAS  \citep[][]{Abey2018Veritas} and HAWC \citep[]{Hona2019, HonaNatAs2021} observations, which showed that TeV spectrum of the Cocoon softens, changing its index from $(-2.1)$ to $\sim(-2.6)$ at the transition energy $\sim 1$ TeV. It is worth mentioning that a few compact clusters (e.g. Westerlund 2, W40) also show the hard-soft behaviour of their $\gamma$-ray spectra  \citep[][]{YangWd2, SunW40}.

Particle acceleration by shock waves and colliding flows in OB-associations was discussed in \citet[]{BT2001, Bykov2001, 2004A&A...424..747P,2010A&A...510A.101F, Bykov2014,2020MNRAS.494.3166V}.
We present here a possible explanation for a piecewise behaviour of the Cygnus Cocoon GeV-TeV spectrum through the modeling of particle acceleration in superbubbles based on the approach of \citet[]{BT2001} and \citet[]{Bykov2001}.
Assuming the hadronic origin of $\gamma$-rays we find the parameters of superbubble needed to provide the break in spectrum and estimate the efficiency of acceleration. 

\section{Proton  acceleration model}

The energy release of massive star clusters allows particle acceleration up to the hundreds of TeV. This can be estimated from the equation connecting the maximum energies of protons, accelerated in the source with the outflows with frozen-in magnetic fields and the magnetic luminosity of the source \citep[see e.g.][]{Lemoine2009}:
\begin{equation}\label{Emax}
E_{\rm max} \approx 6 \times 10^{14} \cdot
\frac{{
\beta^{1/2}_{\rm f}}}{{\rm \Gamma_{\rm f}}}\left(\frac{
{\cal L_M} }{ 10^{35}~\mbox{erg}\,\mbox{s}^{-1}}\right)^{1/2}  \mbox{eV},
\end{equation}
where $\beta_{\rm f} = u_f/c$ is the dimensionless velocity of the flow, $c$ is the speed of light, ${\rm \Gamma}_{\rm f} = 1/\sqrt{1 - \beta_{\rm f}^2 }$. The magnetic luminosity $\cal L_M$ is about a few percent of the mechanical power. For example, the power of the compact cluster  Westerlund 1  is  $\gsim 10^{39}~\mbox{erg}\,\mbox{s}^{-1}$ \citep[see e.g.][]{Muno06}. Cygnus OB2 contains about 120 O stars \citep[][]{2000A&A...360..539K} and it has the mechanical luminosity well above $10^{38}~\mbox{erg}\,\mbox{s}^{-1}$ \citep[see e.g.][]{AharonianNat2019}. 

Most of the star formation activity in Cygnus OB2 occurred between 1 and 7 Myr ago \citep[]{Wright2015}.
Since that time the broad spectrum of magnetohydrodynamic (MHD) fluctuations has formed due to the presence of strong shocks from stellar winds and large-scale flows.
Due to multiple strong shocks, particle propagation in superbubble can be highly intermittent. 
The particle acceleration processes in intermittent systems can be treated by using approximate kinetic equations, averaged over characteristic scales of the system. One characteristic scale is determined by the size of gradient of particle distribution function near a shock wave, while another scale is the energy containing scale of plasma motions, induced by fast winds and strong shocks. 
We use a parameterization for the CR mean free path of particles due to scattering by the quasi-resonant magnetic fluctuations of scales comparable to the CR gyroradius $R_H(p)$ with the CR energy dependence as given by e.g.  \citet{1985crim.book.....T} and  \citet{SMP2007}:
\begin{equation}
\Lambda(p) \simeq l_{\mathrm{corr}}\cdot  \left [\frac{R_H(p)}{l_{\mathrm{corr}}}\right]^{2 - \nu}
\end{equation}   

Here, $\nu$ is the MHD turbulence power-law index. The parameter $l_{ \mathrm {corr}}$, defined so, is of the order of mean distances between strong shocks.


There are two different regimes of CR particle transport in our model depending on the relation of characteristic size of gradient of particle distribution function near a shock wave $l \simeq \kappa(p)/u$  and the effective correlation length $l_{\mathrm{corr}}$. We introduce the momentum $p_{\star}$ where transition occurs: $ {\rm v} \Lambda(p_{\star}) \simeq 3 u l_{ \mathrm {corr}}$, where $\rm v$ is the particle velocity, and $u$ is the amplitude of the bulk plasma speed.
%

There is a strong dependence of the transition momentum $p_{\star}$ on the amplitude of the large-scale turbulent velocity  $u$  as $p_{\star} \propto u^{1/(2-\nu)}$ for $\nu > 3/2$.  From the expressions above for $p_{\star}$ one can get the following estimation for the CR proton energy $\epsilon_{\star}$ at the regime transition point for $\nu =1.7$  
\begin{equation}\label{estar}
\epsilon_{\star} \approx 20\, {\rm GeV}\, \left[\frac{B}{10\, \rm{\mu G}}\right]\cdot \left[\frac{l_{\mathrm{corr}}}{10\, {\rm pc}}\right] \cdot \left[\frac{u}{1,000\, {\rm km ~  s^{-1}}}\right]^{3.33} 
\end{equation}

\subsection{Spectrum of low-energy protons}\label{CRlow}

For low enough energies when $p<p_{\star}$, the CR transport is determined mainly by turbulent advection, particles are tightly connected with plasma motions.  
This regime was studied in non-linear model of \citet{Bykov2001}, where the simplified description of long-wavelength turbulence was applied. The kinetic equation, applicable in intermittent systems with multiple shocks and long-wavelength turbulent motions, was  developed by \citet{Bykov2001} and has the form:
\begin{multline}
    \label{KE:lowE}
      \frac{\partial N}{\partial t} -
       \frac{\partial}{\partial r_{i}} \: \chi_{ij} \:
       \frac{\partial N}{\partial r_{j}}  =
       G  \hat{L} N + \\
     + \frac{1}{p^2} \: \frac{\partial}{\partial p} \: p^4 D \:
      \frac{\partial N}{\partial p} + A {\hat{L}}^2 N +
      2B \hat{L} \hat{P} N + Q(p).
\end{multline}
Here, $N (r, p, t)$ is the CR distribution function, averaged over the ensemble of fluctuations of electric and magnetic fields generated by turbulent plasma motions. The source term $Q(p)$ is determined by injection.
The operators $\hat{L}$ and $\hat{P}$ are given by
\begin{equation}
      \hat{L}= \frac{1}{3p^2} \: \frac{\partial}{\partial p} \:
      p^{3-\gamma} \: \int_{0}^{p} {\rm d}p' \: {p'}^\gamma
      \frac{\partial}{\partial p'} \;;~~~~~
      \hat{P}= \frac{p }{3} \: \frac{\partial}{\partial p} \, .
\end{equation}

Turbulence and shock waves are characterised by the kinetic coefficients $A$, $B$, $D$, $G$ and diffusion tensor $\chi_{ij}$, which describe correlations between shocks and long-wavelength plasma motions.The index $\gamma$ is determined by the shock ensemble properties \citep[][]{bt93}. For the detailed description of Eq. (\ref{KE:lowE}) and definition of these coefficients, see \citet[]{Bykov2014}.
We assume the broad spectral range of magnetic fluctuations $dB_k^2/dk \propto k^{-\nu}$ where the turbulence power-law index $1 \leq \nu \leq 2$.

 For the spectrum calculation the simplified form of Eq. (\ref{KE:lowE}) was used here, where we neglected the cross-correlations coefficients $A$ and  $B$.
It was shown that the efficient conversion of turbulence energy to low-energy particles takes place. 
When $p<p_{\star}$, but injected CRs are relativistic, which is the case for Cygnus Cocoon, proton spectrum demonstrates soft-hard-soft behavior with time, and arrives to asymptotic power-law momentum distribution with the index about 2: $p^2 N(p) \sim p^{-2}$.   

In this context it is important to point out that contrary to the case of Cygnus Cocoon the recent thorough analysis of the multi-wavelength observations  by \citet{2020A&A...635A..96J}  did not reveal any significant departures of the measured $\gamma$-ray emissivity spectrum of the nearby Orion-Eridanus superbubble from the average spectrum measured in the solar neighbourhood. The apparent lack of any $\gamma$-ray emission excess could be understood from Eq.(\ref{estar}) where for the long-wavelength turbulent velocity amplitude below 1,000 $~\mbox{km}\,\mbox{s}^{-1}$ (as it is expected in the  Orion-Eridanus superbubble) $\epsilon_{\star}$ can be well below GeV and the hard CR spectrum may not extend to the proton energies contributing to the photons detected by  {\it Fermi}-LAT. It may indicate that small-scale MHD turbulence filled in the Orion-Eridanus superbubble has a spectral index $\nu \geq 5/3$.      

\subsection{Spectra of high-energy protons}\label{CRhigh}

At high energies  ($p > p_{\star}$) the characteristic diffusion scales of CRs are larger than $l_{\mathrm{corr}}$, which means that particle interacts with several fronts on its free path. 
The particle propagation is described by the Fokker-Planck type equation \citep[cf.][]{2019PhRvD..99h3006L}, which differs from the Eq. (\ref{KE:lowE}) as only the diffusive terms are left. Assuming spherical form of the superbubble: 

\begin{equation}\label{KE:highE}
      \frac{\partial N}{\partial t} -
       \frac{1}{r^2}\frac{\partial}{\partial r} \: r^2 \kappa(p) \:
       \frac{\partial N}{\partial r}  =
      \frac{1}{p^2} \: \frac{\partial}{\partial p} \: p^2 D (p) \:
      \frac{\partial N}{\partial p} ,
\end{equation}
Here, the diffusion in momentum space and spatial diffusion are only left. The spatial diffusion coefficient $\kappa(p) = \rm v \Lambda(p)/3$ and diffusion coefficient in momentum space  $D(p)=p^2/\tau_{\mathrm{acc}}={p^2 u^2}/{9 \kappa(p)}$. 

The steady-state solution of Eq. (\ref{KE:highE}) (i.e. $\partial N/\partial t$=0) depends on momentum like
\begin{equation}
  \label{eq:D}
N(p) = A_0 (p/p_{\star})^{-(\nu +1)/2} K_{\emph a} \left((p/p_{\star})^{2-\nu} \Delta \right) ,   
\end{equation}
where $K_{\emph a}$ is the modified Bessel (Macdonald) function with the index ${\emph a} = (\nu +1)/|4 - 2\nu|$ (for $\nu \neq 2$). Here 
\begin{equation}
\Delta=\frac{\pi}{2} \frac {c \Lambda(p_\star)}{(2-\nu) u R}
\end{equation}
where $R$ is the radius of the system. The detailed derivation of this solution can be found in Appendix A.

\begin{figure}[h]
\centering
\includegraphics [width=85mm] {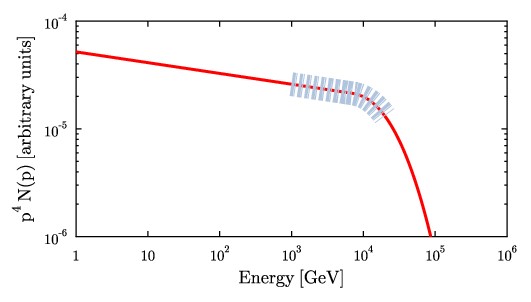}
\caption{The high-energy proton spectrum calculated according to the model described in Section \ref{CRhigh} with the parameters $l_{\mathrm{corr}}$ and $\nu$ determined in Section \ref{GR}. The thick blue line indicates the uncertainties in the  transition region, where the spectrum depends on the matching of the asymptotic solutions.}
\label{SB_CR}
\end{figure}

The resulting proton spectrum in a superbubble is shown in Fig. \ref{SB_CR}. While matching the solutions for low-energy and high-energy regimes, one should have in mind that the transition region around $p_{\star}$ has some uncertainty as the simplified description of the turbulence and CR  propagation is used and 
the solutions are asymptotic.  

\begin{figure}[h]
\centering
\includegraphics [width=85mm] {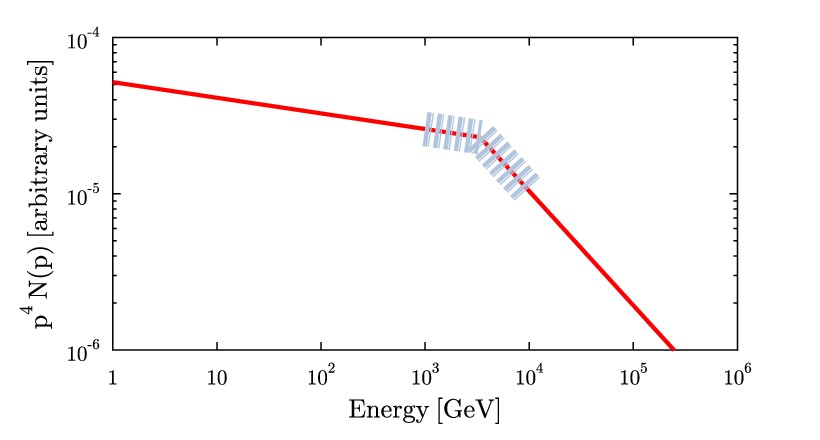}
\caption{The proton spectrum 
derived for the model with the energy independent diffusion coefficient provided by multiple scattering on the ensemble of weak secondary shocks ($\nu =2$) in superbubble  as discussed in Section \ref{CRhigh}. The thick blue line indicates the uncertainties in the transition region, where the spectrum depends on the matching of the asymptotic solutions.}
\label{SB_CR2}
\end{figure}

The CR propagation both in the Galaxy and acceleration sources is governing by their scattering  by MHD turbulence \citep[e.g.][]{1985crim.book.....T,Ptuskin11,PMJSZ2006,ABPW2012,2013ApJ...768...73M,2017PhRvD..95b3007M,Moskalenko:2019MU}. In this study we use simplified models of the diffusive propagation in superbubbles, while the possible non-diffusive regimes are to be studied in a separate paper. The Kolmogorov-type turbulence model of the spectral index $\nu$ = 5/3 is widely used in the global models of CR propagation while the Kraichnan-type ($\nu$ = 3/2) turbulent model is discussed in the context of the low-energy CR reacceleration \citep{SMP2007}. {Strong shocks produced by stellar winds and supernovae could be the dominant local source of the MHD turbulence at some evolution stages of superbubbles and compact stellar clusters.}
It was shown by \citet{BT87,1996ApJ...467..280N} that multiple weak secondary shocks can be produced by interactions of the strong primary shocks provided by the sources of mechanic power with clouds, stellar winds or other types of strong density irregularities. The presence of multiple propagating weak shocks can provide the energy independent CR diffusion coefficient \citep[see][]{BT87} in a rather wide energy range extended to the PeV regime which corresponds to $\nu$ = 2. It is important to point out that such a mechanism gives the power-law proton spectrum as shown in \citet{BT2001} and, therefore, power-law high-energy $\gamma$-ray spectrum as seen by HAWC. For the parameters of the Cygnus Cocoon, to provide the needed power-law index ($\sim -2.6$),
{the diffusion coefficient should} be as small as $ \simeq 3 \times 10^{28}$  cm$^2$ s$^{-1}$ in the PeV regime. The corresponding transition energy should be about $3.4$ TeV. {The alternative high-energy proton spectrum based on these assumptions is shown in Fig. \ref{SB_CR2}. }

\section{Gamma-ray spectrum}\label{GR}
The hadronic mechanism of $\gamma$-ray emission implies proton-proton interactions leading to $\pi^0$ production and subsequent pion-decay radiation. We calculate the emission due to hadronic interactions using parameterizations of \citet{Kelner2006}. For p-p total inelastic cross section we use the most recent parameterization by \citet{KafexhiuATV2014}. Note that Kelner's approach is applicable for proton energies higher than 0.1 TeV, and photon energies higher than 0.1 GeV, which is sufficient for our modeling of GeV-TeV Cocoon spectrum. 

The main parameters of our modeling are 
correlation length $l_{\mathrm{corr}}$, magnetic field $B_0$, and turbulent spectrum index $\nu$. The other parameters are fixed:
association size $R=55$ pc \citep[]{AckermannSB2011,AharonianNat2019}
and the estimate of mean plasma flow velocity $u=$1,500 $~\mbox{km}\,\mbox{s}^{-1}$ based on the measured values of the O- and B- stars wind velocity of 1,000-3,000 $~\mbox{km}\,\mbox{s}^{-1}$ \citep[][]{Seo2018} and estimated supernova remnant shocks velocity which in very rarefied plasma may be $\sim$ 1,500 $~\mbox{km}\,\mbox{s}^{-1}$ for distances $\gsim$ 30 pc  \citep[see e.g.][]{MO77}. While observing SNRs inside superbubbles is not an easy task, the shock velocity estimation of recently discovered SNR G116.6-26.1 in the hot rarefied halo of the Milky Way is consistent with the values discussed above \citep[][]{G116}. 

To fit our model spectrum (Eq. (\ref{eq:D})) to the {\it Fermi}-LAT
 \citep[]{AckermannSB2011}, ARGO \citep[]{Bartoli2014} and HAWC \citep[]{HonaNatAs2021} GeV-TeV data and find corresponding parameters, we use Markov chain Monte Carlo (MCMC) methods based on the Python's {\it emcee} package \citep[]{MCMC}.
Modeling shows that in the range of appropriate values of the mean magnetic field (5-30 $\mu$G), varying the magnetic field doesn't affect much the resulting spectrum, so to prevent extra uncertainty we exclude it from the modeling and fix its value to 15 $\mu$G. For the other two parameters we find the following values with the errors at the 3-sigma confidence level:

\begin{displaymath}
\nu= 1.61^{+0.02}_{-0.02} \\*
\end{displaymath}
\begin{displaymath}
{l_{\mathrm{corr}}}= 27.5^{+11.8}_{-6.1}~ \mathrm{pc}
\end{displaymath}
The fitted $\nu$ value is between the indices for Kolmogorov (5/3) and Kraichnan (3/2) turbulence spectra. Note that for the higher plasma flow velocity $u=$3,000 $~\mbox{km}\,\mbox{s}^{-1}$, we would get the value of $\nu=1.68$ very close to the Kolmogorov spectrum index. 
The $l_{\mathrm{corr}}$, which is in fact the statistical characteristic separation between strong shocks, also has an appropriate value for the $\sim 55$ pc radius of superbubble. For a distance to the source of $1.4$ kpc and gas density of $30$ cm$^{-3}$ \citep[][]{HonaNatAs2021} we obtain the needed proton injection luminosity of the source to be $\sim 1.2 \times 10^{37}~\mbox{erg}\,\mbox{s}^{-1}$ , so the acceleration efficiency is about 0.06 for the  kinetic luminosity of $2 \times 10^{38}~\mbox{erg}\,\mbox{s}^{-1}$  estimated in \citet[][]{AckermannSB2011}.

\begin{figure}[h]
\centering
\includegraphics [width=85mm] {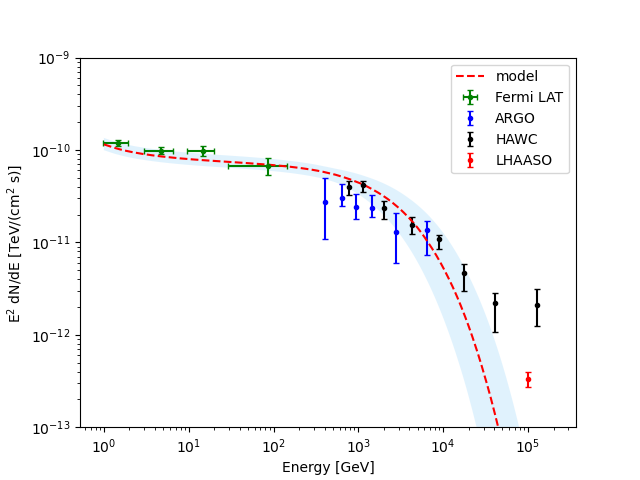}
\caption{The $\gamma$-ray spectral energy distribution for the parameters $\nu=1.61$, $l_{\mathrm{corr}}=27.5$ pc obtained by Markov chain Monte Carlo method. Blue region corresponds to 3-sigma error of the fitting parameters. The extra uncertainty might occur due to the uncertainty of the matching procedure in the parent proton spectrum (see Fig. \ref{SB_CR}).
}
\label{SB_gamma}
\end{figure}
The resulting $\gamma$-ray spectrum with the obtained parameters is shown in Fig. \ref{SB_gamma}.
\begin{figure}[h]
\centering
\includegraphics [width=85mm] {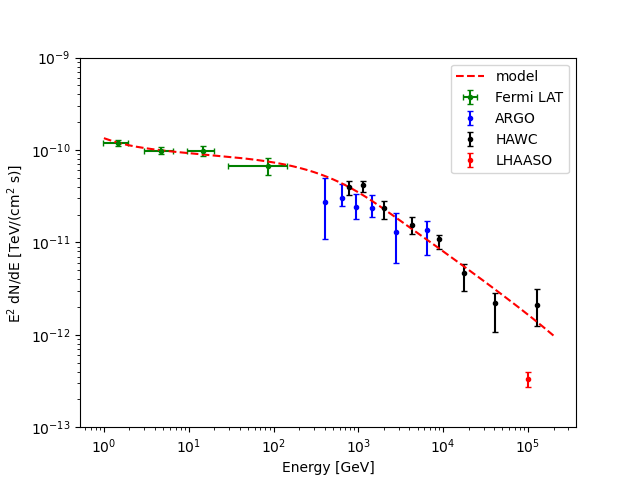}
\caption{The $\gamma$-ray spectral energy distribution for the model where $\nu =2$ and the energy independent diffusion coefficient in superbubble $\kappa \simeq 1.5 \times 10^{28}$  cm$^2$ s$^{-1}$ with the transition energy  $\epsilon_{\star} \simeq 3.4$ TeV. The proton spectrum producing the hadronic $\gamma$-ray emission is shown in Fig. \ref{SB_CR2}.}
\label{SB_gamma2}
\end{figure}
A note of caution is in order here. The transition energy $\epsilon_{\star}$ which matches the two CR transport regimes can not be exactly derived within the model and can only be estimated within a factor of few accuracy.
Therefore the fitting parameters derived with the MCMC 
serve just to illustrate that the theory could reproduce the observed spectrum of the Cygnus Cocoon.

One can see that the hardening of the observed spectrum at the energies higher than 100 TeV is slightly different from our model spectrum.
\citet{HonaNatAs2021} claimed that pulsar wind nebulae (PWNe) powered by known sources (PSR J2021+4026, PSR J2032+4127) do not explain the extended Cocoon emission. A possible time variable behaviour of PSR J2032+4127 in very high energy regime \citep[]{PSR2032_21} could play a role however as well as a yet-undiscovered PWN cannot be excluded as a source of the emission.  Recently, {HAWC} Collaboration investigated similarities in  hardening behaviour in several sources, including Cygnus \citep[]{HAWCHard2021}.
In this respect, we constructed also the $\gamma$-ray spectrum for an alternative  model of CRs propagation in superbubble where the MHD turbulence is dominated by the ensemble of weak shocks (as discussed in Section \ref{CRhigh}) leading to the power-law proton spectrum shown in Fig.\ref{SB_CR2}. The corresponding spectrum of hadronic $\gamma$-ray emission of the system is shown in Fig. \ref{SB_gamma2}. It can successfully match the observed $\gamma$-ray spectrum of the Cygnus Cocoon.          
We also looked at the recent {LHAASO} data from Cygnus region with an approximate flux at $\sim$ 100 TeV provided by \citep[]{LHAASO2021} to see if it matches the Cocoon data and our model. One can see that the only point available so far speaks in favour of our first model shown in Fig. \ref{SB_gamma}. However, the authors cannot claim yet that the source of the $\gamma$-ray emission seen by LHAASO is Cygnus Cocoon. The thorough LHAASO spectral and morphological study of that source is expected in the near future.

 Apart from the extended Cygnus Cocoon there
might be some other particle acceleration sites in the Cygnus region producing PeV-energy protons and neutrinos like that discovered by \citet[]{Dzhappuev2021} which is likely associated with a flaring $\gamma$-ray binary \citep[]{PSR2032_21}. Recently, \citet{Liu2021} suggested that a significant fraction of sub-PeV
gamma rays detected by the Tibet AS+MD array in the galactic disk in the regions 25°$<l<$100° and 50°$<l<$200° may originate
from the source related to Cygnus region. The mentioned above very-high-energy emission detected by LHAASO may also be either of the Cocoon origin, or some other source in the region.
Spectral energy distribution in TeV-PeV region from LHAASO is expected to provide the crucial information about spectral hardening and the need for additional hard-spectrum sources to explain the extended $\gamma$-ray emission in Cygnus region. In the case of the Cygnus Cocoon superbubble one cannot rule out however that some other source(s) with hard spectrum could contribute at the highest energies and account for the last two HAWC data points given in \citet[]{HAWCHard2021}. 

Discoveries of a few interesting ultrahigh-energy $\gamma$-ray sources in some other regions in the Milky Way were reported recently by LHAASO and HAWC. An unidentified extended $\gamma$-ray source LHAASO J0341+5258 located in the Galactic plane has a spectrum fitted with the power-law model of a photon index 2.98 $\pm$ 0.19 in 10 - 200 TeV energy range \citep[]{LHAASO_J0341}. However, the {\it Fermi}-LAT flux upper limit at 10 GeV would require much harder spectrum at lower energies. LHAASO J2108+5157 was significantly detected above 100 TeV with a power-law photon index of 2.83 $\pm$ 0.18 between 20 and 200 TeV while again a harder spectrum is expected at lower energies to be consistent with the existed {\it Fermi}-LAT flux upper limits. The spatial extent of LHAASO J2108+5157 is not yet determined \citep[]{LHAASO_J2108}. The source HAWC J1825-134 which is likely coincident with a giant molecular cloud has a $\gamma$-ray power-law spectrum of rather a hard index 2.28 $\pm$ 0.12 which extends beyond 200 TeV without an apparent cut-off \citep[]{hawc1825}. Dedicated  multi-wavelength studies of these sources are needed to understand whether the model described above is relevant to describe these $\gamma$-ray sources.

\section{Conclusions}
We discuss here a possible explanation for the piecewise behaviour of the Cygnus Cocoon GeV-TeV spectrum with the modeling of particle acceleration and propagation in a superbubble with multiple shocks of different strengths produced by powerful winds of massive stars and supernovae. The softening in the superbubble spectrum at TeV energies in the model indicates a change in the regime of CR propagation inside the accelerator.  

We find fits for the model parameters and show that they have realistic values. The most sensitive parameter in the model is the shape of the spectrum of magnetic fluctuations. If the spectrum of MHD turbulence between the shocks can be approximated by a power-law of index $\nu$ in the broad range of scales resonant to accelerated CRs then the most favorable values of the index are close to $\nu \simeq 5/3$. We find a value for the Cygnus Cocoon $\nu$ = 1.61.

In an alternative model the high-energy CR transport is governed by proton scatterings by an ensemble of multiple secondary weak shocks. The interaction of strong shocks produced by energetic outflows with density inhomogeneities is a source of multiple weak secondary shocks propagating through superbubble. The intermittent distribution of the multiple weak shocks results in energy independent CR diffusion coefficient. This CR propagation model as well provides the good fit to the $\gamma$-ray observations of the Cygnus Cocoon while the spectral shapes of $\gamma$-ray emission are somewhat different.           

The approach presented in this paper may be applied to  other GeV-TeV sources related to associations of young massive stars, as some of compact clusters show similar behaviour in their $\gamma$-ray spectra \citep[][]{YangWd2, SunW40}. In Appendix B we show that the spectrum of the compact star cluster Westerlund 2 can be fitted successfully with our model. However, in some clusters we do not see the significant spectrum softening, e.g., detailed analysis of very-high-energy $\gamma$-ray observations of Westerlund 1 by \citet{AharonianNat2019} revealed a hard $\gamma$-ray spectrum with  a photon index $\sim2.3$ up to $\sim 400$ TeV.  Particle acceleration both in the extended superbubbles and the compact clusters of young massive stars is due to Fermi type mechanism on supersonic MHD flows with multiple shock waves. These type of sources may also have a comparable mechanical power.
Therefore, the differences in their observed $\gamma$-ray spectra are likely determined by the differences in  CR transport within the sources. The large energy density of accelerated CRs in the compact clusters with supernova remnants may result in the dominance of CR driven turbulence providing the Bohm diffusion inside the acceleration region. This may be important for Westerlund 1, explaining the proton spectrum extending up to PeV energies as follows from \citet{AharonianNat2019} observations. Also, the apparent presence of a magnetar in Westerlund 1 indicates its likely interaction with a supernova remnant in the past.

We would like to stress the importance of precise measurements of the $\gamma$-ray spectra in the 100 TeV domain which is crucial to distinguish between the CR propagation regimes and therefore between the  MHD turbulence models in superbubbles discussed above.          

\section*{Acknowledgements}
 The authors would like to thank the anonymous reviewers for their useful comments and suggestions. The authors were supported by the RSF grant 21-72-20020. Some of the modeling was
performed at the ``Tornado'' subsystem of the St.~Petersburg Polytechnic University supercomputing center and at the JSCC RAS.

\appendix
\section{The high-energy proton distribution function}
\begin{equation}
      \frac{\partial N}{\partial t} -
       \frac{1}{r^2}\frac{\partial}{\partial r} \: r^2 \kappa(p) \:
       \frac{\partial N}{\partial r}  =
      \frac{1}{p^2} \: \frac{\partial}{\partial p} \: p^2 D (p) \:
      \frac{\partial N}{\partial p} ,
\end{equation}
The steady state equation is:
\begin{equation}
       \frac{1}{r^2}\frac{\partial}{\partial r} \: r^2 \kappa(p) \:
       \frac{\partial N}{\partial r}  +
      \frac{1}{p^2} \: \frac{\partial}{\partial p} \: p^2 D (p) \:
      \frac{\partial N}{\partial p} =0 
\end{equation}
The diffusion coefficient in momentum space  $D(p)=p^2/\tau_{\mathbf{acc}}={p^2 u^2}/{9 \kappa(p)}$. 
\begin{equation}
       \frac{1}{r^2}\frac{\partial}{\partial r} \: r^2  \:
       \frac{\partial N}{\partial r}  +
      \frac{1}{\kappa (p)} \frac{1}{p^2} \: \frac{\partial}{\partial p} \: p^4 \frac{u^2}{9\kappa(p)} \:
      \frac{\partial N}{\partial p} =0 
\end{equation}
The spatial diffusion coefficient $\kappa(p) = {\rm v} \Lambda(p)/3$, therefore 
\begin{displaymath}
\kappa(p) = \frac{\rm v}{3} \cdot l_{\mathrm{corr}}\cdot  \left [\frac{R_H(p)}{l_{\mathrm{corr}}}\right]^{2 - \nu}.
\end{displaymath}
We are looking for the high-energy solution at $p>p_{\star}$ and in our case these energies are significantly higher than GeV, therefore ${\rm v} \approx c$. Let us introduce the new variable $\eta=p/p_{\star}$, then $\kappa(\eta)=\kappa_0 \eta^{2-\nu}$, where $\kappa_0= (c/3) \cdot l_{\mathrm{corr}} \cdot (c p_{\star}/e B_0 l_{\mathrm{corr}})^{2-\nu}$. For $\eta$ the equation takes the form: 
\begin{equation}
       \frac{1}{r^2}\frac{\partial}{\partial r} \: r^2  \:
       \frac{\partial N}{\partial r}  +
      \frac{1}{\kappa (\eta)} \frac{1}{\eta^2} \: \frac{\partial}{\partial \eta} \: \eta^4 \frac{u^2}{9\kappa(\eta)} \:
      \frac{\partial N}{\partial \eta} =0 
\end{equation}
We separate the variables in the following way:
\begin{displaymath}
N(r,\eta)= f(\eta) \chi(r)/r
\end{displaymath}
For $\chi(r)$ we get:
\begin{equation}
\frac{1}{r} \frac{d}{dr} \left( r^2 \frac{d}{dr} \left (\frac{\chi}{r} \right)\right) =-\lambda \chi,
\end{equation}
and it has the solution
\begin{equation}
\chi = \chi_0 \sin (\sqrt{\lambda} r)    
\end{equation}
where the possible values of $\lambda$ depend on the boundary conditions at the $r=R$ where $R$ is the radius of the system. The main difference between the internal and external regions is assumed to be the absence of significant velocity fluctuations in the external region, i.e. $u=0$, while the spatial diffusion coefficient is close to the internal. Then, after joining the distribution functions and diffusive flows at $r=R$ we obtain
\begin{equation} 
\cos (\sqrt{\lambda} R)=0 \longrightarrow \lambda_n=\left (\frac{\pi}{2R} (2n+1) \right)^2, ~ n=0,1,...  
\end{equation}
As it will be seen below, increasing $n$ leads to the fast fall of the momentum distribution function $f(\eta)$ so we take only one solution with $n=0$. The equation for $f(\eta)$ has the form
\begin{equation}
    \frac{1}{\kappa_0 \eta^{2-\nu}} \cdot \frac{1}{\eta^2} \frac{\partial}{\partial \eta} \eta^4 \frac{u^2}{9 \kappa_0 \eta^{2-\nu}} \frac{\partial f}{\partial \eta}= \lambda f
\end{equation}
We search $f(\eta)$ as 
\begin{displaymath}
f(\eta)=\eta^{-(\nu+1)/2} \Phi(\eta),
\end{displaymath} which leads us to the modified Bessel equation for $\Phi(\eta)$. Throwing away the exponentially growing with energy solution, we obtain:
\begin{equation}
    f(\eta)=A_0 \eta^{-(\nu+1)/2} K_a \left ( \frac{\eta^{2-\nu}3\kappa_0 \sqrt{\lambda_0}}{(2-\nu)u}\right ),
\end{equation}
where $K_{\emph a}$ is the Macdonald function with the index ${\emph a} = (\nu +1)/|4 - 2\nu|$ (for $\nu \neq 2$). With the values of $\kappa_0$ and $\lambda_0$ defined above we get the final particle distribution function in the momentum space (Eq. (\ref{eq:D})). 

\section{The case of Westerlund 2 young massive star cluster}
In order to test our model on a system where the MHD energy source is likely dominated by powerful stellar winds, we apply the approach described in the paper to the compact cluster of young massive stars Westerlund 2. The source shows the similar to the Cygnus Cocoon hard-soft behaviour of its $\gamma$-ray spectrum \citep[][]{YangWd2, Mestre2021} while no apparent supernova remnants neither extended nor compact were reported so far in Westerlund 2. We use the observational data by {\it Fermi} in the GeV region \citep[][]{FermiWd2} and H.E.S.S. in the TeV region \citep[][]{HessWd2}. For high energies we employ the solution obtained in Eq. (\ref{eq:D}), i.e. the model with particle diffusion between strong shocks. The main difference between Cygnus Cocoon and Westerlund 2 is the radius of the cluster.
Although the majority of massive stars are expected to be concentrated in $\leq 1$ pc circle \citep[][]{Alvarez2013}, we take for our calculations the 2 pc radius as the region containing massive stellar winds.
We also need to increase the mean magnetic field $B_0$ to the value $\sim 50$ mkGs, which can be achieved in compact clusters. As before, we take the mean plasma flow velocity as 1,500 $~\mbox{km}\,\mbox{s}^{-1}$.

Within the same approach as before, we get the following results with the errors at the 3-sigma confidence level:

\begin{displaymath}
\nu= 1.53^{+0.07}_{-0.12} \\*
\end{displaymath}
\begin{displaymath}
{l_{\mathrm{corr}}}= 0.51^{+0.45}_{-0.11}~ \mathrm{pc}
\end{displaymath}
We find that parameters' values are reasonable, although both model spectrum and the observed flux values have sizeable errors. Same as for the Cygnus Cocoon, for the higher mean plasma flow velocity (3,000 $~\mbox{km}\,\mbox{s}^{-1}$) we would get turbulence index $\nu \sim 1.63$, which is closer to Kolmogorov index.
\begin{figure}[h]
\centering
\includegraphics [width=85mm] {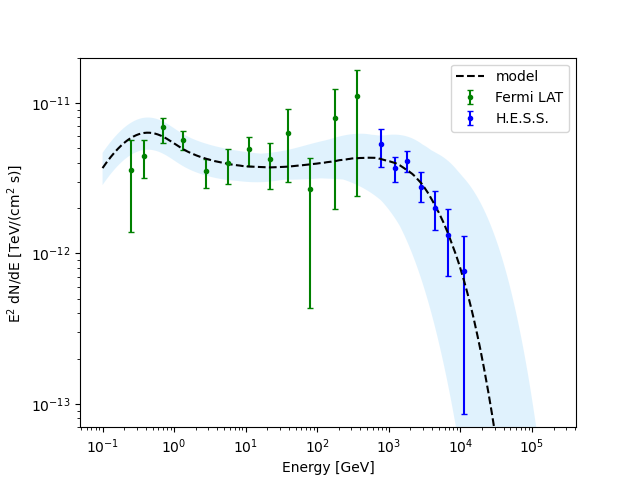}
\caption{The $\gamma$-ray spectral energy distribution for Westerlund 2 compact cluster for the parameters $~ \nu=1.53$, $l_{\mathrm{corr}}=0.51$ pc. 
The model implies the magnetic field $B_0=50$ mkGs and the cluster radius $R=2$ pc. Blue region corresponds to 3-sigma error of the fitting parameters.}
\label{WD2}
\end{figure}
To calculate the CR luminosity we use the distance to the source of 5 kpc and gas density of 25 cm$^{-3}$ \citep[][]{YangWd2} and obtain $\sim 3 \times 10^{36}~\mbox{erg}\,\mbox{s}^{-1}$  which gives the acceleration efficiency of $\sim$ 0.02, considering kinetic luminosity of the cluster $L_k = 2 \times 10^{38}~\mbox{erg}\,\mbox{s}^{-1}$. The resulting $\gamma$-ray spectrum of the compact cluster Westerlund 2 with the parameters defined above is shown in Fig. \ref{WD2}.

We obtain that our model can successfully explain the Westerlund 2 observations, which implies that the suggested mechanism of particle acceleration and propagation can be realized in compact clusters as well as in loose OB-associations. The case of a very interesting source Westerlund 1 which is the most massive young stellar cluster in the Milky Way and, unlike Westerlund 2, has compact supernova remnants, is briefly discussed in Section 4.

 

\bibliographystyle{model5-names}
\biboptions{authoryear}
\bibliography{bib_DCE1}

\begin{thebibliography}{64}
\expandafter\ifx\csname natexlab\endcsname\relax\def\natexlab#1{#1}\fi
\providecommand{\url}[1]{\texttt{#1}}
\providecommand{\href}[2]{#2}
\providecommand{\path}[1]{#1}
\providecommand{\DOIprefix}{doi:}
\providecommand{\ArXivprefix}{arXiv:}
\providecommand{\URLprefix}{URL: }
\providecommand{\Pubmedprefix}{pmid:}
\providecommand{\doi}[1]{\href{http://dx.doi.org/#1}{\path{#1}}}
\providecommand{\Pubmed}[1]{\href{pmid:#1}{\path{#1}}}
\providecommand{\bibinfo}[2]{#2}
\ifx\xfnm\relax \def\xfnm[#1]{\unskip,\space#1}\fi
\bibitem[{{Abdo} et~al.(2012){Abdo}, {Abeysekara}, {Allen}, {Aune}, {Berley},
  {Bonamente}, {Christopher}, {DeYoung}, {Dingus}, {Ellsworth},
  {Galbraith-Frew}, {Gonzalez}, {Goodman}, {Hoffman}, {H{\"u}ntemeyer}, {Hui},
  {Kolterman}, {Linnemann}, {McEnery}, {Mincer}, {Morgan}, {Nemethy}, {Pretz},
  {Ryan}, {Saz Parkinson}, {Shoup}, {Sinnis}, {Smith}, {Vasileiou}, {Walker},
  {Williams} \& {Yodh}}]{Abdo2012}
\bibinfo{author}{{Abdo}, A.~A.}, \bibinfo{author}{{Abeysekara}, U.},
  \bibinfo{author}{{Allen}, B.~T.}, \bibinfo{author}{{Aune}, T.},
  \bibinfo{author}{{Berley}, D.}, \bibinfo{author}{{Bonamente}, E.},
  \bibinfo{author}{{Christopher}, G.~E.}, \bibinfo{author}{{DeYoung}, T.},
  \bibinfo{author}{{Dingus}, B.~L.}, \bibinfo{author}{{Ellsworth}, R.~W.},
  \bibinfo{author}{{Galbraith-Frew}, J.~G.}, \bibinfo{author}{{Gonzalez},
  M.~M.}, \bibinfo{author}{{Goodman}, J.~A.}, \bibinfo{author}{{Hoffman},
  C.~M.}, \bibinfo{author}{{H{\"u}ntemeyer}, P.~H.}, \bibinfo{author}{{Hui},
  C.~M.}, \bibinfo{author}{{Kolterman}, B.~E.}, \bibinfo{author}{{Linnemann},
  J.~T.}, \bibinfo{author}{{McEnery}, J.~E.}, \bibinfo{author}{{Mincer},
  A.~I.}, \bibinfo{author}{{Morgan}, T.}, \bibinfo{author}{{Nemethy}, P.},
  \bibinfo{author}{{Pretz}, J.}, \bibinfo{author}{{Ryan}, J.~M.},
  \bibinfo{author}{{Saz Parkinson}, P.~M.}, \bibinfo{author}{{Shoup}, A.},
  \bibinfo{author}{{Sinnis}, G.}, \bibinfo{author}{{Smith}, A.~J.},
  \bibinfo{author}{{Vasileiou}, V.}, \bibinfo{author}{{Walker}, G.~P.},
  \bibinfo{author}{{Williams}, D.~A.}, \& \bibinfo{author}{{Yodh}, G.~B.}
  (\bibinfo{year}{2012}).
\newblock \bibinfo{title}{{Spectrum and Morphology of the Two Brightest Milagro
  Sources in the Cygnus Region: MGRO J2019+37 and MGRO J2031+41}}.
\newblock {\it \bibinfo{journal}{The Astrophysical Journal}\/},  {\it
  \bibinfo{volume}{753}\/}\bibinfo{issue}{(2)}, \bibinfo{pages}{159}.
  \DOIprefix\doi{10.1088/0004-637X/753/2/159}.
  \href{http://arxiv.org/abs/1202.0846}{\tt arXiv:1202.0846}.
\bibitem[{{Abdo} et~al.(2007{\natexlab{a}}){Abdo}, {Allen}, {Berley},
  {Blaufuss}, {Casanova}, {Chen}, {Coyne}, {Delay}, {Dingus}, {Ellsworth},
  {Fleysher}, {Fleysher}, {Gebauer}, {Gonzalez}, {Goodman}, {Hays}, {Hoffman},
  {Kolterman}, {Kelley}, {Lansdell}, {Linnemann}, {McEnery}, {Mincer},
  {Moskalenko}, {Nemethy}, {Noyes}, {Ryan}, {Samuelson}, {Saz Parkinson},
  {Schneider}, {Shoup}, {Sinnis}, {Smith}, {Strong}, {Sullivan}, {Vasileiou},
  {Walker}, {Williams}, {Xu} \& {Yodh}}]{Abdo2007a}
\bibinfo{author}{{Abdo}, A.~A.}, \bibinfo{author}{{Allen}, B.},
  \bibinfo{author}{{Berley}, D.}, \bibinfo{author}{{Blaufuss}, E.},
  \bibinfo{author}{{Casanova}, S.}, \bibinfo{author}{{Chen}, C.},
  \bibinfo{author}{{Coyne}, D.~G.}, \bibinfo{author}{{Delay}, R.~S.},
  \bibinfo{author}{{Dingus}, B.~L.}, \bibinfo{author}{{Ellsworth}, R.~W.},
  \bibinfo{author}{{Fleysher}, L.}, \bibinfo{author}{{Fleysher}, R.},
  \bibinfo{author}{{Gebauer}, I.}, \bibinfo{author}{{Gonzalez}, M.~M.},
  \bibinfo{author}{{Goodman}, J.~A.}, \bibinfo{author}{{Hays}, E.},
  \bibinfo{author}{{Hoffman}, C.~M.}, \bibinfo{author}{{Kolterman}, B.~E.},
  \bibinfo{author}{{Kelley}, L.~A.}, \bibinfo{author}{{Lansdell}, C.~P.},
  \bibinfo{author}{{Linnemann}, J.~T.}, \bibinfo{author}{{McEnery}, J.~E.},
  \bibinfo{author}{{Mincer}, A.~I.}, \bibinfo{author}{{Moskalenko}, I.~V.},
  \bibinfo{author}{{Nemethy}, P.}, \bibinfo{author}{{Noyes}, D.},
  \bibinfo{author}{{Ryan}, J.~M.}, \bibinfo{author}{{Samuelson}, F.~W.},
  \bibinfo{author}{{Saz Parkinson}, P.~M.}, \bibinfo{author}{{Schneider}, M.},
  \bibinfo{author}{{Shoup}, A.}, \bibinfo{author}{{Sinnis}, G.},
  \bibinfo{author}{{Smith}, A.~J.}, \bibinfo{author}{{Strong}, A.~W.},
  \bibinfo{author}{{Sullivan}, G.~W.}, \bibinfo{author}{{Vasileiou}, V.},
  \bibinfo{author}{{Walker}, G.~P.}, \bibinfo{author}{{Williams}, D.~A.},
  \bibinfo{author}{{Xu}, X.~W.}, \& \bibinfo{author}{{Yodh}, G.~B.}
  (\bibinfo{year}{2007}{\natexlab{a}}).
\newblock \bibinfo{title}{{Discovery of TeV Gamma-Ray Emission from the Cygnus
  Region of the Galaxy}}.
\newblock {\it \bibinfo{journal}{The Astrophysical Journal Letters}\/},  {\it
  \bibinfo{volume}{658}\/}\bibinfo{issue}{(1)}, \bibinfo{pages}{L33--L36}.
  \DOIprefix\doi{10.1086/513696}.
  \href{http://arxiv.org/abs/astro-ph/0611691}{\tt arXiv:astro-ph/0611691}.
\bibitem[{{Abdo} et~al.(2007{\natexlab{b}}){Abdo}, {Allen}, {Berley},
  {Casanova}, {Chen}, {Coyne}, {Dingus}, {Ellsworth}, {Fleysher}, {Fleysher},
  {Gonzalez}, {Goodman}, {Hays}, {Hoffman}, {Hopper}, {H{\"u}ntemeyer},
  {Kolterman}, {Lansdell}, {Linnemann}, {McEnery}, {Mincer}, {Nemethy},
  {Noyes}, {Ryan}, {Saz Parkinson}, {Shoup}, {Sinnis}, {Smith}, {Sullivan},
  {Vasileiou}, {Walker}, {Williams}, {Xu} \& {Yodh}}]{Abdo2007b}
\bibinfo{author}{{Abdo}, A.~A.}, \bibinfo{author}{{Allen}, B.},
  \bibinfo{author}{{Berley}, D.}, \bibinfo{author}{{Casanova}, S.},
  \bibinfo{author}{{Chen}, C.}, \bibinfo{author}{{Coyne}, D.~G.},
  \bibinfo{author}{{Dingus}, B.~L.}, \bibinfo{author}{{Ellsworth}, R.~W.},
  \bibinfo{author}{{Fleysher}, L.}, \bibinfo{author}{{Fleysher}, R.},
  \bibinfo{author}{{Gonzalez}, M.~M.}, \bibinfo{author}{{Goodman}, J.~A.},
  \bibinfo{author}{{Hays}, E.}, \bibinfo{author}{{Hoffman}, C.~M.},
  \bibinfo{author}{{Hopper}, B.}, \bibinfo{author}{{H{\"u}ntemeyer}, P.~H.},
  \bibinfo{author}{{Kolterman}, B.~E.}, \bibinfo{author}{{Lansdell}, C.~P.},
  \bibinfo{author}{{Linnemann}, J.~T.}, \bibinfo{author}{{McEnery}, J.~E.},
  \bibinfo{author}{{Mincer}, A.~I.}, \bibinfo{author}{{Nemethy}, P.},
  \bibinfo{author}{{Noyes}, D.}, \bibinfo{author}{{Ryan}, J.~M.},
  \bibinfo{author}{{Saz Parkinson}, P.~M.}, \bibinfo{author}{{Shoup}, A.},
  \bibinfo{author}{{Sinnis}, G.}, \bibinfo{author}{{Smith}, A.~J.},
  \bibinfo{author}{{Sullivan}, G.~W.}, \bibinfo{author}{{Vasileiou}, V.},
  \bibinfo{author}{{Walker}, G.~P.}, \bibinfo{author}{{Williams}, D.~A.},
  \bibinfo{author}{{Xu}, X.~W.}, \& \bibinfo{author}{{Yodh}, G.~B.}
  (\bibinfo{year}{2007}{\natexlab{b}}).
\newblock \bibinfo{title}{{TeV Gamma-Ray Sources from a Survey of the Galactic
  Plane with Milagro}}.
\newblock {\it \bibinfo{journal}{The Astrophysical Journal Letters}\/},  {\it
  \bibinfo{volume}{664}\/}\bibinfo{issue}{(2)}, \bibinfo{pages}{L91--L94}.
  \DOIprefix\doi{10.1086/520717}. \href{http://arxiv.org/abs/0705.0707}{\tt
  arXiv:0705.0707}.
\bibitem[{{Abeysekara} et~al.(2021){Abeysekara}, {Albert}, {Alfaro}, {Alvarez},
  {Camacho}, {Arteaga-Vel{\'a}zquez}, {Arunbabu}, {Rojas}, {Solares},
  {Baghmanyan}, {Belmont-Moreno}, {BenZvi}, {Blandford}, {Brisbois},
  {Caballero-Mora}, {Capistr{\'a}n}, {Carrami{\~n}ana}, {Casanova}, {Cotti},
  {Le{\'o}n}, {De la Fuente}, {Hernandez}, {Dingus}, {DuVernois}, {Durocher},
  {D{\'\i}az-V{\'e}lez}, {Ellsworth}, {Engel}, {Espinoza}, {Fan}, {Fang},
  {Fleischhack}, {Fraija}, {Galv{\'a}n-G{\'a}mez}, {Garcia},
  {Garc{\'\i}a-Gonz{\'a}lez}, {Garfias}, {Giacinti}, {Gonz{\'a}lez}, {Goodman},
  {Harding}, {Hernandez}, {Hinton}, {Hona}, {Huang}, {Hueyotl-Zahuantitla},
  {H{\"u}ntemeyer}, {Iriarte}, {Jardin-Blicq}, {Joshi}, {Kieda}, {Lara}, {Lee},
  {Vargas}, {Linnemann}, {Longinotti}, {Luis-Raya}, {Lundeen}, {Malone},
  {Martinez}, {Martinez-Castellanos}, {Mart{\'\i}nez-Castro}, {Matthews},
  {Miranda-Romagnoli}, {Morales-Soto}, {Moreno}, {Mostaf{\'a}}, {Nayerhoda},
  {Nellen}, {Newbold}, {Nisa}, {Noriega-Papaqui}, {Olivera-Nieto}, {Omodei},
  {Peisker}, {P{\'e}rez Araujo}, {P{\'e}rez-P{\'e}rez}, {Ren}, {Rho},
  {Rosa-Gonz{\'a}lez}, {Ruiz-Velasco}, {Salazar}, {Greus}, {Sandoval},
  {Schneider}, {Schoorlemmer}, {Serna}, {Smith}, {Springer}, {Surajbali},
  {Tollefson}, {Torres}, {Torres-Escobedo}, {Ure{\~n}a-Mena}, {Weisgarber},
  {Werner}, {Willox}, {Zepeda}, {Zhou} et~al.}]{HonaNatAs2021}
\bibinfo{author}{{Abeysekara}, A.~U.}, \bibinfo{author}{{Albert}, A.},
  \bibinfo{author}{{Alfaro}, R.}, \bibinfo{author}{{Alvarez}, C.},
  \bibinfo{author}{{Camacho}, J.~R.~A.},
  \bibinfo{author}{{Arteaga-Vel{\'a}zquez}, J.~C.},
  \bibinfo{author}{{Arunbabu}, K.~P.}, \bibinfo{author}{{Rojas}, D.~A.},
  \bibinfo{author}{{Solares}, H.~A.~A.}, \bibinfo{author}{{Baghmanyan}, V.},
  \bibinfo{author}{{Belmont-Moreno}, E.}, \bibinfo{author}{{BenZvi}, S.~Y.},
  \bibinfo{author}{{Blandford}, R.}, \bibinfo{author}{{Brisbois}, C.},
  \bibinfo{author}{{Caballero-Mora}, K.~S.}, \bibinfo{author}{{Capistr{\'a}n},
  T.}, \bibinfo{author}{{Carrami{\~n}ana}, A.}, \bibinfo{author}{{Casanova},
  S.}, \bibinfo{author}{{Cotti}, U.}, \bibinfo{author}{{Le{\'o}n}, S. C.~d.},
  \bibinfo{author}{{De la Fuente}, E.}, \bibinfo{author}{{Hernandez}, R.~D.},
  \bibinfo{author}{{Dingus}, B.~L.}, \bibinfo{author}{{DuVernois}, M.~A.},
  \bibinfo{author}{{Durocher}, M.}, \bibinfo{author}{{D{\'\i}az-V{\'e}lez},
  J.~C.}, \bibinfo{author}{{Ellsworth}, R.~W.}, \bibinfo{author}{{Engel}, K.},
  \bibinfo{author}{{Espinoza}, C.}, \bibinfo{author}{{Fan}, K.~L.},
  \bibinfo{author}{{Fang}, K.}, \bibinfo{author}{{Fleischhack}, H.},
  \bibinfo{author}{{Fraija}, N.}, \bibinfo{author}{{Galv{\'a}n-G{\'a}mez}, A.},
  \bibinfo{author}{{Garcia}, D.}, \bibinfo{author}{{Garc{\'\i}a-Gonz{\'a}lez},
  J.~A.}, \bibinfo{author}{{Garfias}, F.}, \bibinfo{author}{{Giacinti}, G.},
  \bibinfo{author}{{Gonz{\'a}lez}, M.~M.}, \bibinfo{author}{{Goodman}, J.~A.},
  \bibinfo{author}{{Harding}, J.~P.}, \bibinfo{author}{{Hernandez}, S.},
  \bibinfo{author}{{Hinton}, J.}, \bibinfo{author}{{Hona}, B.},
  \bibinfo{author}{{Huang}, D.}, \bibinfo{author}{{Hueyotl-Zahuantitla}, F.},
  \bibinfo{author}{{H{\"u}ntemeyer}, P.}, \bibinfo{author}{{Iriarte}, A.},
  \bibinfo{author}{{Jardin-Blicq}, A.}, \bibinfo{author}{{Joshi}, V.},
  \bibinfo{author}{{Kieda}, D.}, \bibinfo{author}{{Lara}, A.},
  \bibinfo{author}{{Lee}, W.~H.}, \bibinfo{author}{{Vargas}, H.~L.},
  \bibinfo{author}{{Linnemann}, J.~T.}, \bibinfo{author}{{Longinotti}, A.~L.},
  \bibinfo{author}{{Luis-Raya}, G.}, \bibinfo{author}{{Lundeen}, J.},
  \bibinfo{author}{{Malone}, K.}, \bibinfo{author}{{Martinez}, O.},
  \bibinfo{author}{{Martinez-Castellanos}, I.},
  \bibinfo{author}{{Mart{\'\i}nez-Castro}, J.}, \bibinfo{author}{{Matthews},
  J.~A.}, \bibinfo{author}{{Miranda-Romagnoli}, P.},
  \bibinfo{author}{{Morales-Soto}, J.~A.}, \bibinfo{author}{{Moreno}, E.},
  \bibinfo{author}{{Mostaf{\'a}}, M.}, \bibinfo{author}{{Nayerhoda}, A.},
  \bibinfo{author}{{Nellen}, L.}, \bibinfo{author}{{Newbold}, M.},
  \bibinfo{author}{{Nisa}, M.~U.}, \bibinfo{author}{{Noriega-Papaqui}, R.},
  \bibinfo{author}{{Olivera-Nieto}, L.}, \bibinfo{author}{{Omodei}, N.},
  \bibinfo{author}{{Peisker}, A.}, \bibinfo{author}{{P{\'e}rez Araujo}, Y.},
  \bibinfo{author}{{P{\'e}rez-P{\'e}rez}, E.~G.}, \bibinfo{author}{{Ren}, Z.},
  \bibinfo{author}{{Rho}, C.~D.}, \bibinfo{author}{{Rosa-Gonz{\'a}lez}, D.},
  \bibinfo{author}{{Ruiz-Velasco}, E.}, \bibinfo{author}{{Salazar}, H.},
  \bibinfo{author}{{Greus}, F.~S.}, \bibinfo{author}{{Sandoval}, A.},
  \bibinfo{author}{{Schneider}, M.}, \bibinfo{author}{{Schoorlemmer}, H.},
  \bibinfo{author}{{Serna}, F.}, \bibinfo{author}{{Smith}, A.~J.},
  \bibinfo{author}{{Springer}, R.~W.}, \bibinfo{author}{{Surajbali}, P.},
  \bibinfo{author}{{Tollefson}, K.}, \bibinfo{author}{{Torres}, I.},
  \bibinfo{author}{{Torres-Escobedo}, R.}, \bibinfo{author}{{Ure{\~n}a-Mena},
  F.}, \bibinfo{author}{{Weisgarber}, T.}, \bibinfo{author}{{Werner}, F.},
  \bibinfo{author}{{Willox}, E.}, \bibinfo{author}{{Zepeda}, A.},
  \bibinfo{author}{{Zhou}, H.} et~al. (\bibinfo{year}{2021}).
\newblock \bibinfo{title}{{HAWC observations of the acceleration of
  very-high-energy cosmic rays in the Cygnus Cocoon}}.
\newblock {\it \bibinfo{journal}{Nature Astronomy}\/},  {\it
  \bibinfo{volume}{5}\/}, \bibinfo{pages}{465--471}.
  \DOIprefix\doi{10.1038/s41550-021-01318-y}.
  \href{http://arxiv.org/abs/2103.06820}{\tt arXiv:2103.06820}.
\bibitem[{Abeysekara et~al.(2021)Abeysekara, Albert, Alfaro, Alvarez, Álvarez
  Romero, Angeles~Camacho, Arteaga~Velazquez, Kollamparambil, Avila~Rojas,
  Ayala~Solares, Babu, Baghmanyan, Barber, Becerra~Gonzalez, Belmont-Moreno,
  BenZvi, Berley, Brisbois, Caballero~Mora, Capistrán, Carramiñana, Casanova,
  Chaparro-Amaro, Cotti, Cotzomi, Coutiño~de Leon, de~la Fuente, de~León,
  Diaz, Diaz~Hernandez, Díaz~Vélez, Dingus, Durocher, DuVernois, Ellsworth,
  Engel, Espinoza~Hernández, Fan, Fang, Fernandez~Alonso, Fick, Fleischhack,
  Flores, Fraija, Garcia~Aguilar, Garcia-Gonzalez, García-Luna,
  García-Torales, Garfias, Giacinti, Goksu, González, Goodman, Harding,
  Hernández~Cadena, Herzog, Hinton, Hona, Huang, Hueyotl-Zahuantitla, Hui,
  Humensky, Hüntemeyer, Iriarte, Jardin-Blicq, Jhee, Joshi, Kieda, Kunde,
  Kunwar, Lara, Lee, Lee, Lennarz, Leon~Vargas, Linnemann, Longinotti,
  Lopez-Coto, Luis-Raya, Lundeen, Malone, Marandon, Martinez,
  Martinez~Castellanos, Martínez~Huerta, Martínez-Castro, Matthews, McEnery,
  Miranda-Romagnoli, Morales~Soto, Moreno~Barbosa, Mostafa, Nayerhoda, Nellen,
  Newbold, Nisa, Noriega-Papaqui, Olivera-Nieto, Omodei et~al.}]{HAWCHard2021}
\bibinfo{author}{Abeysekara, A.~U.}, \bibinfo{author}{Albert, A.},
  \bibinfo{author}{Alfaro, R.}, \bibinfo{author}{Alvarez, C.},
  \bibinfo{author}{Álvarez Romero, J. d.~D.},
  \bibinfo{author}{Angeles~Camacho, J.~R.}, \bibinfo{author}{Arteaga~Velazquez,
  J.~C.}, \bibinfo{author}{Kollamparambil, A.~B.},
  \bibinfo{author}{Avila~Rojas, D.~O.}, \bibinfo{author}{Ayala~Solares, H.~A.},
  \bibinfo{author}{Babu, R.}, \bibinfo{author}{Baghmanyan, V.},
  \bibinfo{author}{Barber, A.~S.}, \bibinfo{author}{Becerra~Gonzalez, J.},
  \bibinfo{author}{Belmont-Moreno, E.}, \bibinfo{author}{BenZvi, S.},
  \bibinfo{author}{Berley, D.}, \bibinfo{author}{Brisbois, C.},
  \bibinfo{author}{Caballero~Mora, K.~S.}, \bibinfo{author}{Capistrán, T.},
  \bibinfo{author}{Carramiñana, A.}, \bibinfo{author}{Casanova, S.},
  \bibinfo{author}{Chaparro-Amaro, O.}, \bibinfo{author}{Cotti, U.},
  \bibinfo{author}{Cotzomi, J.}, \bibinfo{author}{Coutiño~de Leon, S.},
  \bibinfo{author}{de~la Fuente, E.}, \bibinfo{author}{de~León, C.~L.},
  \bibinfo{author}{Diaz, L.}, \bibinfo{author}{Diaz~Hernandez, R.},
  \bibinfo{author}{Díaz~Vélez, J.~C.}, \bibinfo{author}{Dingus, B.},
  \bibinfo{author}{Durocher, M.}, \bibinfo{author}{DuVernois, M.},
  \bibinfo{author}{Ellsworth, R.}, \bibinfo{author}{Engel, K.},
  \bibinfo{author}{Espinoza~Hernández, M.~C.}, \bibinfo{author}{Fan, J.},
  \bibinfo{author}{Fang, K.}, \bibinfo{author}{Fernandez~Alonso, M.},
  \bibinfo{author}{Fick, B.}, \bibinfo{author}{Fleischhack, H.},
  \bibinfo{author}{Flores, J.~L.}, \bibinfo{author}{Fraija, N.~I.},
  \bibinfo{author}{Garcia~Aguilar, D.}, \bibinfo{author}{Garcia-Gonzalez,
  J.~A.}, \bibinfo{author}{García-Luna, J.~L.},
  \bibinfo{author}{García-Torales, G.}, \bibinfo{author}{Garfias, F.},
  \bibinfo{author}{Giacinti, G.}, \bibinfo{author}{Goksu, H.},
  \bibinfo{author}{González, M.~M.}, \bibinfo{author}{Goodman, J.~A.},
  \bibinfo{author}{Harding, J.~P.}, \bibinfo{author}{Hernández~Cadena, S.},
  \bibinfo{author}{Herzog, I.}, \bibinfo{author}{Hinton, J.},
  \bibinfo{author}{Hona, B.}, \bibinfo{author}{Huang, D.},
  \bibinfo{author}{Hueyotl-Zahuantitla, F.}, \bibinfo{author}{Hui, M.},
  \bibinfo{author}{Humensky, B.}, \bibinfo{author}{Hüntemeyer, P.},
  \bibinfo{author}{Iriarte, A.}, \bibinfo{author}{Jardin-Blicq, A.},
  \bibinfo{author}{Jhee, H.}, \bibinfo{author}{Joshi, V.},
  \bibinfo{author}{Kieda, D.}, \bibinfo{author}{Kunde, G.~J.},
  \bibinfo{author}{Kunwar, S.}, \bibinfo{author}{Lara, A.},
  \bibinfo{author}{Lee, J.}, \bibinfo{author}{Lee, W.~H.},
  \bibinfo{author}{Lennarz, D.}, \bibinfo{author}{Leon~Vargas, H.},
  \bibinfo{author}{Linnemann, J.}, \bibinfo{author}{Longinotti, A.~L.},
  \bibinfo{author}{Lopez-Coto, R.}, \bibinfo{author}{Luis-Raya, G.},
  \bibinfo{author}{Lundeen, J.}, \bibinfo{author}{Malone, K.},
  \bibinfo{author}{Marandon, V.}, \bibinfo{author}{Martinez, O.},
  \bibinfo{author}{Martinez~Castellanos, I.},
  \bibinfo{author}{Martínez~Huerta, H.}, \bibinfo{author}{Martínez-Castro,
  J.}, \bibinfo{author}{Matthews, J.}, \bibinfo{author}{McEnery, J.},
  \bibinfo{author}{Miranda-Romagnoli, P.}, \bibinfo{author}{Morales~Soto,
  J.~A.}, \bibinfo{author}{Moreno~Barbosa, E.}, \bibinfo{author}{Mostafa, M.},
  \bibinfo{author}{Nayerhoda, A.}, \bibinfo{author}{Nellen, L.},
  \bibinfo{author}{Newbold, M.}, \bibinfo{author}{Nisa, M.~U.},
  \bibinfo{author}{Noriega-Papaqui, R.}, \bibinfo{author}{Olivera-Nieto, L.},
  \bibinfo{author}{Omodei, N.} et~al. (\bibinfo{year}{2021}).
\newblock \bibinfo{title}{{A search for spectral hardening in HAWC sources
  above 56 TeV}}.
\newblock {\it \bibinfo{journal}{Proceedings of Science}\/},  {\it
  \bibinfo{volume}{ICRC2021}\/}, \bibinfo{pages}{811}.
  \DOIprefix\doi{10.22323/1.395.0811}.
\bibitem[{{Abeysekara} et~al.(2018){Abeysekara}, {Archer}, {Aune}, {Benbow},
  {Bird}, {Brose}, {Buchovecky}, {Bugaev}, {Cui}, {Daniel}, {Falcone}, {Feng},
  {Finley}, {Fleischhack}, {Flinders}, {Fortson}, {Furniss}, {Gotthelf},
  {Grube}, {Hanna}, {Hervet}, {Holder}, {Huang}, {Hughes}, {Humensky},
  {H{\"u}tten}, {Johnson}, {Kaaret}, {Kar}, {Kelley-Hoskins}, {Kertzman},
  {Kieda}, {Krause}, {Kumar}, {Lang}, {Lin}, {Maier}, {McArthur}, {Moriarty},
  {Mukherjee}, {O'Brien}, {Ong}, {Otte}, {Pandel}, {Park}, {Petrashyk}, {Pohl},
  {Popkow}, {Pueschel}, {Quinn}, {Ragan}, {Reynolds}, {Richards}, {Roache},
  {Rousselle}, {Rulten}, {Sadeh}, {Santander}, {Sembroski}, {Shahinyan},
  {Tyler}, {Vassiliev}, {Wakely}, {Ward}, {Weinstein}, {Wells}, {Wilcox},
  {Wilhelm}, {Williams} \& {Zitzer}}]{Abey2018Veritas}
\bibinfo{author}{{Abeysekara}, A.~U.}, \bibinfo{author}{{Archer}, A.},
  \bibinfo{author}{{Aune}, T.}, \bibinfo{author}{{Benbow}, W.},
  \bibinfo{author}{{Bird}, R.}, \bibinfo{author}{{Brose}, R.},
  \bibinfo{author}{{Buchovecky}, M.}, \bibinfo{author}{{Bugaev}, V.},
  \bibinfo{author}{{Cui}, W.}, \bibinfo{author}{{Daniel}, M.~K.},
  \bibinfo{author}{{Falcone}, A.}, \bibinfo{author}{{Feng}, Q.},
  \bibinfo{author}{{Finley}, J.~P.}, \bibinfo{author}{{Fleischhack}, H.},
  \bibinfo{author}{{Flinders}, A.}, \bibinfo{author}{{Fortson}, L.},
  \bibinfo{author}{{Furniss}, A.}, \bibinfo{author}{{Gotthelf}, E.~V.},
  \bibinfo{author}{{Grube}, J.}, \bibinfo{author}{{Hanna}, D.},
  \bibinfo{author}{{Hervet}, O.}, \bibinfo{author}{{Holder}, J.},
  \bibinfo{author}{{Huang}, K.}, \bibinfo{author}{{Hughes}, G.},
  \bibinfo{author}{{Humensky}, T.~B.}, \bibinfo{author}{{H{\"u}tten}, M.},
  \bibinfo{author}{{Johnson}, C.~A.}, \bibinfo{author}{{Kaaret}, P.},
  \bibinfo{author}{{Kar}, P.}, \bibinfo{author}{{Kelley-Hoskins}, N.},
  \bibinfo{author}{{Kertzman}, M.}, \bibinfo{author}{{Kieda}, D.},
  \bibinfo{author}{{Krause}, M.}, \bibinfo{author}{{Kumar}, S.},
  \bibinfo{author}{{Lang}, M.~J.}, \bibinfo{author}{{Lin}, T.~T.~Y.},
  \bibinfo{author}{{Maier}, G.}, \bibinfo{author}{{McArthur}, S.},
  \bibinfo{author}{{Moriarty}, P.}, \bibinfo{author}{{Mukherjee}, R.},
  \bibinfo{author}{{O'Brien}, S.}, \bibinfo{author}{{Ong}, R.~A.},
  \bibinfo{author}{{Otte}, A.~N.}, \bibinfo{author}{{Pandel}, D.},
  \bibinfo{author}{{Park}, N.}, \bibinfo{author}{{Petrashyk}, A.},
  \bibinfo{author}{{Pohl}, M.}, \bibinfo{author}{{Popkow}, A.},
  \bibinfo{author}{{Pueschel}, E.}, \bibinfo{author}{{Quinn}, J.},
  \bibinfo{author}{{Ragan}, K.}, \bibinfo{author}{{Reynolds}, P.~T.},
  \bibinfo{author}{{Richards}, G.~T.}, \bibinfo{author}{{Roache}, E.},
  \bibinfo{author}{{Rousselle}, J.}, \bibinfo{author}{{Rulten}, C.},
  \bibinfo{author}{{Sadeh}, I.}, \bibinfo{author}{{Santander}, M.},
  \bibinfo{author}{{Sembroski}, G.~H.}, \bibinfo{author}{{Shahinyan}, K.},
  \bibinfo{author}{{Tyler}, J.}, \bibinfo{author}{{Vassiliev}, V.~V.},
  \bibinfo{author}{{Wakely}, S.~P.}, \bibinfo{author}{{Ward}, J.~E.},
  \bibinfo{author}{{Weinstein}, A.}, \bibinfo{author}{{Wells}, R.~M.},
  \bibinfo{author}{{Wilcox}, P.}, \bibinfo{author}{{Wilhelm}, A.},
  \bibinfo{author}{{Williams}, D.~A.}, \& \bibinfo{author}{{Zitzer}, B.}
  (\bibinfo{year}{2018}).
\newblock \bibinfo{title}{{A Very High Energy {\ensuremath{\gamma}}-Ray Survey
  toward the Cygnus Region of the Galaxy}}.
\newblock {\it \bibinfo{journal}{The Astrophysical Journal}\/},  {\it
  \bibinfo{volume}{861}\/}\bibinfo{issue}{(2)}, \bibinfo{pages}{134}.
  \DOIprefix\doi{10.3847/1538-4357/aac4a2}.
  \href{http://arxiv.org/abs/1805.05989}{\tt arXiv:1805.05989}.
\bibitem[{{Ackermann} et~al.(2011){Ackermann}, {Ajello}, {Allafort}, {Baldini},
  {Ballet}, {Barbiellini}, {Bastieri}, {Belfiore}, {Bellazzini}, {Berenji},
  {Blandford}, {Bloom}, {Bonamente}, {Borgland}, {Bottacini}, {Brigida},
  {Bruel}, {Buehler}, {Buson}, {Caliandro}, {Cameron}, {Caraveo}, {Casandjian},
  {Cecchi}, {Chekhtman}, {Cheung}, {Chiang}, {Ciprini}, {Claus},
  {Cohen-Tanugi}, {de Angelis}, {de Palma}, {Dermer}, {do Couto e Silva},
  {Drell}, {Dumora}, {Favuzzi}, {Fegan}, {Focke}, {Fortin}, {Fukazawa},
  {Fusco}, {Gargano}, {Germani}, {Giglietto}, {Giordano}, {Giroletti},
  {Glanzman}, {Godfrey}, {Grenier}, {Guillemot}, {Guiriec}, {Hadasch},
  {Hanabata}, {Okumura}, {Orlando}, {Ormes}, {Ozaki}, {Paneque}, {Parent},
  {Pesce-Rollins}, {Pierbattista}, {Piron}, {Pohl}, {Prokhorov}, {Rain{\`o}},
  {Rando}, {Razzano}, {Reposeur}, {Ritz}, {Parkinson}, {Sgr{\`o}}, {Siskind},
  {Smith}, {Spinelli}, {Strong}, {Takahashi}, {Tanaka}, {Thayer}, {Thayer},
  {Thompson}, {Tibaldo}, {Torres}, {Tosti}, {Tramacere}, {Troja}, {Uchiyama},
  {Vandenbroucke}, {Vasileiou}, {Vianello}, {Vitale}, {Waite}, {Wang}, {Winer},
  {Wood}, {Yang}, {Zimmer} \& {Bontemps}}]{AckermannSB2011}
\bibinfo{author}{{Ackermann}, M.}, \bibinfo{author}{{Ajello}, M.},
  \bibinfo{author}{{Allafort}, A.}, \bibinfo{author}{{Baldini}, L.},
  \bibinfo{author}{{Ballet}, J.}, \bibinfo{author}{{Barbiellini}, G.},
  \bibinfo{author}{{Bastieri}, D.}, \bibinfo{author}{{Belfiore}, A.},
  \bibinfo{author}{{Bellazzini}, R.}, \bibinfo{author}{{Berenji}, B.},
  \bibinfo{author}{{Blandford}, R.~D.}, \bibinfo{author}{{Bloom}, E.~D.},
  \bibinfo{author}{{Bonamente}, E.}, \bibinfo{author}{{Borgland}, A.~W.},
  \bibinfo{author}{{Bottacini}, E.}, \bibinfo{author}{{Brigida}, M.},
  \bibinfo{author}{{Bruel}, P.}, \bibinfo{author}{{Buehler}, R.},
  \bibinfo{author}{{Buson}, S.}, \bibinfo{author}{{Caliandro}, G.~A.},
  \bibinfo{author}{{Cameron}, R.~A.}, \bibinfo{author}{{Caraveo}, P.~A.},
  \bibinfo{author}{{Casandjian}, J.~M.}, \bibinfo{author}{{Cecchi}, C.},
  \bibinfo{author}{{Chekhtman}, A.}, \bibinfo{author}{{Cheung}, C.~C.},
  \bibinfo{author}{{Chiang}, J.}, \bibinfo{author}{{Ciprini}, S.},
  \bibinfo{author}{{Claus}, R.}, \bibinfo{author}{{Cohen-Tanugi}, J.},
  \bibinfo{author}{{de Angelis}, A.}, \bibinfo{author}{{de Palma}, F.},
  \bibinfo{author}{{Dermer}, C.~D.}, \bibinfo{author}{{do Couto e Silva}, E.},
  \bibinfo{author}{{Drell}, P.~S.}, \bibinfo{author}{{Dumora}, D.},
  \bibinfo{author}{{Favuzzi}, C.}, \bibinfo{author}{{Fegan}, S.~J.},
  \bibinfo{author}{{Focke}, W.~B.}, \bibinfo{author}{{Fortin}, P.},
  \bibinfo{author}{{Fukazawa}, Y.}, \bibinfo{author}{{Fusco}, P.},
  \bibinfo{author}{{Gargano}, F.}, \bibinfo{author}{{Germani}, S.},
  \bibinfo{author}{{Giglietto}, N.}, \bibinfo{author}{{Giordano}, F.},
  \bibinfo{author}{{Giroletti}, M.}, \bibinfo{author}{{Glanzman}, T.},
  \bibinfo{author}{{Godfrey}, G.}, \bibinfo{author}{{Grenier}, I.~A.},
  \bibinfo{author}{{Guillemot}, L.}, \bibinfo{author}{{Guiriec}, S.},
  \bibinfo{author}{{Hadasch}, D.}, \bibinfo{author}{{Hanabata}, Y.},
  \bibinfo{author}{{Okumura}, A.}, \bibinfo{author}{{Orlando}, E.},
  \bibinfo{author}{{Ormes}, J.~F.}, \bibinfo{author}{{Ozaki}, M.},
  \bibinfo{author}{{Paneque}, D.}, \bibinfo{author}{{Parent}, D.},
  \bibinfo{author}{{Pesce-Rollins}, M.}, \bibinfo{author}{{Pierbattista}, M.},
  \bibinfo{author}{{Piron}, F.}, \bibinfo{author}{{Pohl}, M.},
  \bibinfo{author}{{Prokhorov}, D.}, \bibinfo{author}{{Rain{\`o}}, S.},
  \bibinfo{author}{{Rando}, R.}, \bibinfo{author}{{Razzano}, M.},
  \bibinfo{author}{{Reposeur}, T.}, \bibinfo{author}{{Ritz}, S.},
  \bibinfo{author}{{Parkinson}, P.~M.~S.}, \bibinfo{author}{{Sgr{\`o}}, C.},
  \bibinfo{author}{{Siskind}, E.~J.}, \bibinfo{author}{{Smith}, P.~D.},
  \bibinfo{author}{{Spinelli}, P.}, \bibinfo{author}{{Strong}, A.~W.},
  \bibinfo{author}{{Takahashi}, H.}, \bibinfo{author}{{Tanaka}, T.},
  \bibinfo{author}{{Thayer}, J.~G.}, \bibinfo{author}{{Thayer}, J.~B.},
  \bibinfo{author}{{Thompson}, D.~J.}, \bibinfo{author}{{Tibaldo}, L.},
  \bibinfo{author}{{Torres}, D.~F.}, \bibinfo{author}{{Tosti}, G.},
  \bibinfo{author}{{Tramacere}, A.}, \bibinfo{author}{{Troja}, E.},
  \bibinfo{author}{{Uchiyama}, Y.}, \bibinfo{author}{{Vandenbroucke}, J.},
  \bibinfo{author}{{Vasileiou}, V.}, \bibinfo{author}{{Vianello}, G.},
  \bibinfo{author}{{Vitale}, V.}, \bibinfo{author}{{Waite}, A.~P.},
  \bibinfo{author}{{Wang}, P.}, \bibinfo{author}{{Winer}, B.~L.},
  \bibinfo{author}{{Wood}, K.~S.}, \bibinfo{author}{{Yang}, Z.},
  \bibinfo{author}{{Zimmer}, S.}, \& \bibinfo{author}{{Bontemps}, S.}
  (\bibinfo{year}{2011}).
\newblock \bibinfo{title}{{A Cocoon of Freshly Accelerated Cosmic Rays Detected
  by Fermi in the Cygnus Superbubble}}.
\newblock {\it \bibinfo{journal}{Science}\/},  {\it \bibinfo{volume}{334}\/},
  \bibinfo{pages}{1103--}. \DOIprefix\doi{10.1126/science.1210311}.
\bibitem[{{Ackermann} et~al.(2017){Ackermann}, {Ajello}, {Baldini}, {Ballet},
  {Barbiellini}, {Bastieri}, {Bellazzini}, {Bissaldi}, {Bloom}, {Bonino},
  {Bottacini}, {Brandt}, {Bregeon}, {Bruel}, {Buehler}, {Cameron}, {Caragiulo},
  {Caraveo}, {Castro}, {Cavazzuti}, {Cecchi}, {Charles}, {Chekhtman}, {Cheung},
  {Chiaro}, {Ciprini}, {Cohen}, {Costantin}, {Costanza}, {Cutini}, {D'Ammando},
  {de Palma}, {Desiante}, {Digel}, {Di Lalla}, {Di Mauro}, {Di Venere},
  {Favuzzi}, {Fegan}, {Ferrara}, {Franckowiak}, {Fukazawa}, {Funk}, {Fusco},
  {Gargano}, {Gasparrini}, {Giglietto}, {Giordano}, {Giroletti}, {Green},
  {Grenier}, {Grondin}, {Guillemot}, {Guiriec}, {Harding}, {Hays}, {Hewitt},
  {Horan}, {Hou}, {J{\'o}hannesson}, {Kamae}, {Kuss}, {La Mura}, {Larsson},
  {Lemoine-Goumard}, {Li}, {Longo}, {Loparco}, {Lubrano}, {Magill}, {Maldera},
  {Malyshev}, {Manfreda}, {Mazziotta}, {Michelson}, {Mitthumsiri}, {Mizuno},
  {Monzani}, {Morselli}, {Moskalenko}, {Negro}, {Nuss}, {Ohsugi}, {Omodei},
  {Orienti}, {Orlando}, {Ormes}, {Paliya}, {Paneque}, {Perkins}, {Persic},
  {Pesce-Rollins}, {Petrosian}, {Piron}, {Porter}, {Principe}, {Rain{\`o}},
  {Rando}, {Razzano} et~al.}]{FermiWd2}
\bibinfo{author}{{Ackermann}, M.}, \bibinfo{author}{{Ajello}, M.},
  \bibinfo{author}{{Baldini}, L.}, \bibinfo{author}{{Ballet}, J.},
  \bibinfo{author}{{Barbiellini}, G.}, \bibinfo{author}{{Bastieri}, D.},
  \bibinfo{author}{{Bellazzini}, R.}, \bibinfo{author}{{Bissaldi}, E.},
  \bibinfo{author}{{Bloom}, E.~D.}, \bibinfo{author}{{Bonino}, R.},
  \bibinfo{author}{{Bottacini}, E.}, \bibinfo{author}{{Brandt}, T.~J.},
  \bibinfo{author}{{Bregeon}, J.}, \bibinfo{author}{{Bruel}, P.},
  \bibinfo{author}{{Buehler}, R.}, \bibinfo{author}{{Cameron}, R.~A.},
  \bibinfo{author}{{Caragiulo}, M.}, \bibinfo{author}{{Caraveo}, P.~A.},
  \bibinfo{author}{{Castro}, D.}, \bibinfo{author}{{Cavazzuti}, E.},
  \bibinfo{author}{{Cecchi}, C.}, \bibinfo{author}{{Charles}, E.},
  \bibinfo{author}{{Chekhtman}, A.}, \bibinfo{author}{{Cheung}, C.~C.},
  \bibinfo{author}{{Chiaro}, G.}, \bibinfo{author}{{Ciprini}, S.},
  \bibinfo{author}{{Cohen}, J.~M.}, \bibinfo{author}{{Costantin}, D.},
  \bibinfo{author}{{Costanza}, F.}, \bibinfo{author}{{Cutini}, S.},
  \bibinfo{author}{{D'Ammando}, F.}, \bibinfo{author}{{de Palma}, F.},
  \bibinfo{author}{{Desiante}, R.}, \bibinfo{author}{{Digel}, S.~W.},
  \bibinfo{author}{{Di Lalla}, N.}, \bibinfo{author}{{Di Mauro}, M.},
  \bibinfo{author}{{Di Venere}, L.}, \bibinfo{author}{{Favuzzi}, C.},
  \bibinfo{author}{{Fegan}, S.~J.}, \bibinfo{author}{{Ferrara}, E.~C.},
  \bibinfo{author}{{Franckowiak}, A.}, \bibinfo{author}{{Fukazawa}, Y.},
  \bibinfo{author}{{Funk}, S.}, \bibinfo{author}{{Fusco}, P.},
  \bibinfo{author}{{Gargano}, F.}, \bibinfo{author}{{Gasparrini}, D.},
  \bibinfo{author}{{Giglietto}, N.}, \bibinfo{author}{{Giordano}, F.},
  \bibinfo{author}{{Giroletti}, M.}, \bibinfo{author}{{Green}, D.},
  \bibinfo{author}{{Grenier}, I.~A.}, \bibinfo{author}{{Grondin}, M.~H.},
  \bibinfo{author}{{Guillemot}, L.}, \bibinfo{author}{{Guiriec}, S.},
  \bibinfo{author}{{Harding}, A.~K.}, \bibinfo{author}{{Hays}, E.},
  \bibinfo{author}{{Hewitt}, J.~W.}, \bibinfo{author}{{Horan}, D.},
  \bibinfo{author}{{Hou}, X.}, \bibinfo{author}{{J{\'o}hannesson}, G.},
  \bibinfo{author}{{Kamae}, T.}, \bibinfo{author}{{Kuss}, M.},
  \bibinfo{author}{{La Mura}, G.}, \bibinfo{author}{{Larsson}, S.},
  \bibinfo{author}{{Lemoine-Goumard}, M.}, \bibinfo{author}{{Li}, J.},
  \bibinfo{author}{{Longo}, F.}, \bibinfo{author}{{Loparco}, F.},
  \bibinfo{author}{{Lubrano}, P.}, \bibinfo{author}{{Magill}, J.~D.},
  \bibinfo{author}{{Maldera}, S.}, \bibinfo{author}{{Malyshev}, D.},
  \bibinfo{author}{{Manfreda}, A.}, \bibinfo{author}{{Mazziotta}, M.~N.},
  \bibinfo{author}{{Michelson}, P.~F.}, \bibinfo{author}{{Mitthumsiri}, W.},
  \bibinfo{author}{{Mizuno}, T.}, \bibinfo{author}{{Monzani}, M.~E.},
  \bibinfo{author}{{Morselli}, A.}, \bibinfo{author}{{Moskalenko}, I.~V.},
  \bibinfo{author}{{Negro}, M.}, \bibinfo{author}{{Nuss}, E.},
  \bibinfo{author}{{Ohsugi}, T.}, \bibinfo{author}{{Omodei}, N.},
  \bibinfo{author}{{Orienti}, M.}, \bibinfo{author}{{Orlando}, E.},
  \bibinfo{author}{{Ormes}, J.~F.}, \bibinfo{author}{{Paliya}, V.~S.},
  \bibinfo{author}{{Paneque}, D.}, \bibinfo{author}{{Perkins}, J.~S.},
  \bibinfo{author}{{Persic}, M.}, \bibinfo{author}{{Pesce-Rollins}, M.},
  \bibinfo{author}{{Petrosian}, V.}, \bibinfo{author}{{Piron}, F.},
  \bibinfo{author}{{Porter}, T.~A.}, \bibinfo{author}{{Principe}, G.},
  \bibinfo{author}{{Rain{\`o}}, S.}, \bibinfo{author}{{Rando}, R.},
  \bibinfo{author}{{Razzano}, M.} et~al. (\bibinfo{year}{2017}).
\newblock \bibinfo{title}{{Search for Extended Sources in the Galactic Plane
  Using Six Years of Fermi-Large Area Telescope Pass 8 Data above 10 GeV}}.
\newblock {\it \bibinfo{journal}{The Astrophysical Journal}\/},  {\it
  \bibinfo{volume}{843}\/}\bibinfo{issue}{(2)}, \bibinfo{pages}{139}.
  \DOIprefix\doi{10.3847/1538-4357/aa775a}.
  \href{http://arxiv.org/abs/1702.00476}{\tt arXiv:1702.00476}.
\bibitem[{{Aharonian} et~al.(2002){Aharonian}, {Akhperjanian}, {Beilicke},
  {Bernl{\"o}hr}, {B{\"o}rst}, {Bojahr}, {Bolz}, {Coarasa}, {Contreras},
  {Cortina}, {Denninghoff}, {Fonseca}, {Girma}, {G{\"o}tting}, {Heinzelmann},
  {Hermann}, {Heusler}, {Hofmann}, {Horns}, {Jung}, {Kankanyan}, {Kestel},
  {Kettler}, {Kohnle}, {Konopelko}, {Kornmeyer}, {Kranich}, {Krawczynski},
  {Lampeitl}, {Lopez}, {Lorenz}, {Lucarelli}, {Magnussen}, {Mang}, {Meyer},
  {Milite}, {Mirzoyan}, {Moralejo}, {Ona}, {Panter}, {Plyasheshnikov}, {Prahl},
  {P{\"u}hlhofer}, {Rauterberg}, {Reyes}, {Rhode}, {Ripken}, {R{\"o}hring},
  {Rowell}, {Sahakian}, {Samorski}, {Schilling}, {Schr{\"o}der}, {Siems},
  {Sobzynska}, {Stamm}, {Tluczykont}, {V{\"o}lk}, {Wiedner}, {Wittek},
  {Uchiyama}, {Takahashi} \& {HEGRA Collaboration}}]{2002A&A...393L..37A}
\bibinfo{author}{{Aharonian}, F.}, \bibinfo{author}{{Akhperjanian}, A.},
  \bibinfo{author}{{Beilicke}, M.}, \bibinfo{author}{{Bernl{\"o}hr}, K.},
  \bibinfo{author}{{B{\"o}rst}, H.}, \bibinfo{author}{{Bojahr}, H.},
  \bibinfo{author}{{Bolz}, O.}, \bibinfo{author}{{Coarasa}, T.},
  \bibinfo{author}{{Contreras}, J.}, \bibinfo{author}{{Cortina}, J.},
  \bibinfo{author}{{Denninghoff}, S.}, \bibinfo{author}{{Fonseca}, V.},
  \bibinfo{author}{{Girma}, M.}, \bibinfo{author}{{G{\"o}tting}, N.},
  \bibinfo{author}{{Heinzelmann}, G.}, \bibinfo{author}{{Hermann}, G.},
  \bibinfo{author}{{Heusler}, A.}, \bibinfo{author}{{Hofmann}, W.},
  \bibinfo{author}{{Horns}, D.}, \bibinfo{author}{{Jung}, I.},
  \bibinfo{author}{{Kankanyan}, R.}, \bibinfo{author}{{Kestel}, M.},
  \bibinfo{author}{{Kettler}, J.}, \bibinfo{author}{{Kohnle}, A.},
  \bibinfo{author}{{Konopelko}, A.}, \bibinfo{author}{{Kornmeyer}, H.},
  \bibinfo{author}{{Kranich}, D.}, \bibinfo{author}{{Krawczynski}, H.},
  \bibinfo{author}{{Lampeitl}, H.}, \bibinfo{author}{{Lopez}, M.},
  \bibinfo{author}{{Lorenz}, E.}, \bibinfo{author}{{Lucarelli}, F.},
  \bibinfo{author}{{Magnussen}, N.}, \bibinfo{author}{{Mang}, O.},
  \bibinfo{author}{{Meyer}, H.}, \bibinfo{author}{{Milite}, M.},
  \bibinfo{author}{{Mirzoyan}, R.}, \bibinfo{author}{{Moralejo}, A.},
  \bibinfo{author}{{Ona}, E.}, \bibinfo{author}{{Panter}, M.},
  \bibinfo{author}{{Plyasheshnikov}, A.}, \bibinfo{author}{{Prahl}, J.},
  \bibinfo{author}{{P{\"u}hlhofer}, G.}, \bibinfo{author}{{Rauterberg}, G.},
  \bibinfo{author}{{Reyes}, R.}, \bibinfo{author}{{Rhode}, W.},
  \bibinfo{author}{{Ripken}, J.}, \bibinfo{author}{{R{\"o}hring}, A.},
  \bibinfo{author}{{Rowell}, G.~P.}, \bibinfo{author}{{Sahakian}, V.},
  \bibinfo{author}{{Samorski}, M.}, \bibinfo{author}{{Schilling}, M.},
  \bibinfo{author}{{Schr{\"o}der}, F.}, \bibinfo{author}{{Siems}, M.},
  \bibinfo{author}{{Sobzynska}, D.}, \bibinfo{author}{{Stamm}, W.},
  \bibinfo{author}{{Tluczykont}, M.}, \bibinfo{author}{{V{\"o}lk}, H.~J.},
  \bibinfo{author}{{Wiedner}, C.~A.}, \bibinfo{author}{{Wittek}, W.},
  \bibinfo{author}{{Uchiyama}, Y.}, \bibinfo{author}{{Takahashi}, T.}, \&
  \bibinfo{author}{{HEGRA Collaboration}} (\bibinfo{year}{2002}).
\newblock \bibinfo{title}{{An unidentified TeV source in the vicinity of Cygnus
  OB2}}.
\newblock {\it \bibinfo{journal}{Astronomy \& Astrophysics}\/},  {\it
  \bibinfo{volume}{393}\/}, \bibinfo{pages}{L37--L40}.
  \DOIprefix\doi{10.1051/0004-6361:20021171}.
  \href{http://arxiv.org/abs/astro-ph/0207528}{\tt arXiv:astro-ph/0207528}.
\bibitem[{{Aharonian} et~al.(2012){Aharonian}, {Bykov}, {Parizot}, {Ptuskin} \&
  {Watson}}]{ABPW2012}
\bibinfo{author}{{Aharonian}, F.}, \bibinfo{author}{{Bykov}, A.},
  \bibinfo{author}{{Parizot}, E.}, \bibinfo{author}{{Ptuskin}, V.}, \&
  \bibinfo{author}{{Watson}, A.} (\bibinfo{year}{2012}).
\newblock \bibinfo{title}{{Cosmic Rays in Galactic and Extragalactic Magnetic
  Fields}}.
\newblock {\it \bibinfo{journal}{Space Science Reviews}\/},  {\it
  \bibinfo{volume}{166}\/}, \bibinfo{pages}{97--132}.
  \href{http://arxiv.org/abs/1105.0131}{\tt arXiv:1105.0131}.
\bibitem[{{Aharonian} et~al.(2019){Aharonian}, {Yang} \& {de O{\~n}a
  Wilhelmi}}]{AharonianNat2019}
\bibinfo{author}{{Aharonian}, F.}, \bibinfo{author}{{Yang}, R.}, \&
  \bibinfo{author}{{de O{\~n}a Wilhelmi}, E.} (\bibinfo{year}{2019}).
\newblock \bibinfo{title}{{Massive stars as major factories of Galactic cosmic
  rays}}.
\newblock {\it \bibinfo{journal}{Nature Astronomy}\/},  {\it
  \bibinfo{volume}{3}\/}, \bibinfo{pages}{561--567}.
  \DOIprefix\doi{10.1038/s41550-019-0724-0}.
  \href{http://arxiv.org/abs/1804.02331}{\tt arXiv:1804.02331}.
\bibitem[{{Albert} et~al.(2021){Albert}, {Alfaro}, {Alvarez}, {Camacho},
  {Arteaga-Vel{\'a}zquez}, {Arunbabu}, {Rojas}, {Ayala Solares}, {Baghmanyan},
  {Belmont-Moreno}, {BenZvi}, {Brisbois}, {Capistr{\'a}n}, {Carrami{\~n}ana},
  {Casanova}, {Cotti}, {Cotzomi}, {Fuente}, {Hernandez}, {Dingus}, {DuVernois},
  {Durocher}, {D{\'\i}az-V{\'e}lez}, {Engel}, {Espinoza}, {Fang},
  {Fleischhack}, {Fraija}, {Galv{\'a}n-G{\'a}mez}, {Garcia},
  {Garc{\'\i}a-Gonz{\'a}lez}, {Garfias}, {Giacinti}, {Gonz{\'a}lez}, {Goodman},
  {Harding}, {Hona}, {Huang}, {Hueyotl-Zahuantitla}, {H{\"u}ntemeyer},
  {Iriarte}, {Jardin-Blicq}, {Joshi}, {Kunde}, {Lara}, {Lee}, {Vargas},
  {Linnemann}, {Longinotti}, {Luis-Raya}, {Lundeen}, {Malone}, {Marandon},
  {Martinez}, {Mart{\'\i}nez-Castro}, {Matthews}, {Miranda-Romagnoli},
  {Moreno}, {Mostaf{\'a}}, {Nayerhoda}, {Nellen}, {Newbold}, {Nisa},
  {Noriega-Papaqui}, {Omodei}, {Peisker}, {Araujo}, {P{\'e}rez-P{\'e}rez},
  {Rho}, {Rosa-Gonz{\'a}lez}, {Salazar}, {Greus}, {Sandoval}, {Schneider},
  {Serna}, {Springer}, {Tollefson}, {Torres}, {Torres-Escobedo},
  {Ure{\~n}a-Mena}, {Villase{\~n}or}, {Willox}, {Zhou} \&
  {Le{\'o}n}}]{hawc1825}
\bibinfo{author}{{Albert}, A.}, \bibinfo{author}{{Alfaro}, R.},
  \bibinfo{author}{{Alvarez}, C.}, \bibinfo{author}{{Camacho}, J.~R.~A.},
  \bibinfo{author}{{Arteaga-Vel{\'a}zquez}, J.~C.},
  \bibinfo{author}{{Arunbabu}, K.~P.}, \bibinfo{author}{{Rojas}, D.~A.},
  \bibinfo{author}{{Ayala Solares}, H.~A.}, \bibinfo{author}{{Baghmanyan}, V.},
  \bibinfo{author}{{Belmont-Moreno}, E.}, \bibinfo{author}{{BenZvi}, S.~Y.},
  \bibinfo{author}{{Brisbois}, C.}, \bibinfo{author}{{Capistr{\'a}n}, T.},
  \bibinfo{author}{{Carrami{\~n}ana}, A.}, \bibinfo{author}{{Casanova}, S.},
  \bibinfo{author}{{Cotti}, U.}, \bibinfo{author}{{Cotzomi}, J.},
  \bibinfo{author}{{Fuente}, E. D.~l.}, \bibinfo{author}{{Hernandez}, R.~D.},
  \bibinfo{author}{{Dingus}, B.~L.}, \bibinfo{author}{{DuVernois}, M.~A.},
  \bibinfo{author}{{Durocher}, M.}, \bibinfo{author}{{D{\'\i}az-V{\'e}lez},
  J.~C.}, \bibinfo{author}{{Engel}, K.}, \bibinfo{author}{{Espinoza}, C.},
  \bibinfo{author}{{Fang}, K.}, \bibinfo{author}{{Fleischhack}, H.},
  \bibinfo{author}{{Fraija}, N.}, \bibinfo{author}{{Galv{\'a}n-G{\'a}mez}, A.},
  \bibinfo{author}{{Garcia}, D.}, \bibinfo{author}{{Garc{\'\i}a-Gonz{\'a}lez},
  J.~A.}, \bibinfo{author}{{Garfias}, F.}, \bibinfo{author}{{Giacinti}, G.},
  \bibinfo{author}{{Gonz{\'a}lez}, M.~M.}, \bibinfo{author}{{Goodman}, J.~A.},
  \bibinfo{author}{{Harding}, J.~P.}, \bibinfo{author}{{Hona}, B.},
  \bibinfo{author}{{Huang}, D.}, \bibinfo{author}{{Hueyotl-Zahuantitla}, F.},
  \bibinfo{author}{{H{\"u}ntemeyer}, P.}, \bibinfo{author}{{Iriarte}, A.},
  \bibinfo{author}{{Jardin-Blicq}, A.}, \bibinfo{author}{{Joshi}, V.},
  \bibinfo{author}{{Kunde}, G.~J.}, \bibinfo{author}{{Lara}, A.},
  \bibinfo{author}{{Lee}, W.~H.}, \bibinfo{author}{{Vargas}, H.~L.},
  \bibinfo{author}{{Linnemann}, J.~T.}, \bibinfo{author}{{Longinotti}, A.~L.},
  \bibinfo{author}{{Luis-Raya}, G.}, \bibinfo{author}{{Lundeen}, J.},
  \bibinfo{author}{{Malone}, K.}, \bibinfo{author}{{Marandon}, V.},
  \bibinfo{author}{{Martinez}, O.}, \bibinfo{author}{{Mart{\'\i}nez-Castro},
  J.}, \bibinfo{author}{{Matthews}, J.~A.},
  \bibinfo{author}{{Miranda-Romagnoli}, P.}, \bibinfo{author}{{Moreno}, E.},
  \bibinfo{author}{{Mostaf{\'a}}, M.}, \bibinfo{author}{{Nayerhoda}, A.},
  \bibinfo{author}{{Nellen}, L.}, \bibinfo{author}{{Newbold}, M.},
  \bibinfo{author}{{Nisa}, M.~U.}, \bibinfo{author}{{Noriega-Papaqui}, R.},
  \bibinfo{author}{{Omodei}, N.}, \bibinfo{author}{{Peisker}, A.},
  \bibinfo{author}{{Araujo}, Y.~P.}, \bibinfo{author}{{P{\'e}rez-P{\'e}rez},
  E.~G.}, \bibinfo{author}{{Rho}, C.~D.}, \bibinfo{author}{{Rosa-Gonz{\'a}lez},
  D.}, \bibinfo{author}{{Salazar}, H.}, \bibinfo{author}{{Greus}, F.~S.},
  \bibinfo{author}{{Sandoval}, A.}, \bibinfo{author}{{Schneider}, M.},
  \bibinfo{author}{{Serna}, F.}, \bibinfo{author}{{Springer}, R.~W.},
  \bibinfo{author}{{Tollefson}, K.}, \bibinfo{author}{{Torres}, I.},
  \bibinfo{author}{{Torres-Escobedo}, R.}, \bibinfo{author}{{Ure{\~n}a-Mena},
  F.}, \bibinfo{author}{{Villase{\~n}or}, L.}, \bibinfo{author}{{Willox}, E.},
  \bibinfo{author}{{Zhou}, H.}, \& \bibinfo{author}{{Le{\'o}n}, C.~d.}
  (\bibinfo{year}{2021}).
\newblock \bibinfo{title}{{Evidence of 200 TeV Photons from HAWC J1825-134}}.
\newblock {\it \bibinfo{journal}{The Astrophysical Journal Letters}\/},  {\it
  \bibinfo{volume}{907}\/}\bibinfo{issue}{(2)}, \bibinfo{pages}{L30}.
  \DOIprefix\doi{10.3847/2041-8213/abd77b}.
  \href{http://arxiv.org/abs/2012.15275}{\tt arXiv:2012.15275}.
\bibitem[{{Bartoli} et~al.(2014){Bartoli}, {Bernardini}, {Bi}, {Branchini},
  {Budano}, {Camarri}, {Cao}, {Cardarelli}, {Catalanotti}, {Chen}, {Chen},
  {Creti}, {Cui}, {Dai}, {D'Amone}, {Danzengluobu}, {De Mitri}, {D'Ettorre
  Piazzoli}, {Di Girolamo}, {Di Sciascio}, {Feng}, {Feng}, {Feng}, {Gou},
  {Guo}, {He}, {Hu}, {Hu}, {Iacovacci}, {Iuppa}, {Jia}, {Labaciren}, {Li},
  {Liguori}, {Liu}, {Liu}, {Liu}, {Lu}, {Ma}, {Ma}, {Mancarella}, {Mari},
  {Marsella}, {Martello}, {Mastroianni}, {Montini}, {Ning}, {Panareo},
  {Perrone}, {Pistilli}, {Ruggieri}, {Salvini}, {Santonico}, {Shen}, {Sheng},
  {Shi}, {Surdo}, {Tan}, {Vallania}, {Vernetto}, {Vigorito}, {Wang}, {Wu},
  {Wu}, {Xue}, {Yang}, {Yang}, {Yao}, {Yuan}, {Zha}, {Zhang}, {Zhang}, {Zhang},
  {Zhang}, {Zhao}, {Zhaxiciren}, {Zhaxisangzhu}, {Zhou}, {Zhu}, {Zhu}, {Zizzi}
  \& {ARGO-YBJ Collaboration}}]{Bartoli2014}
\bibinfo{author}{{Bartoli}, B.}, \bibinfo{author}{{Bernardini}, P.},
  \bibinfo{author}{{Bi}, X.~J.}, \bibinfo{author}{{Branchini}, P.},
  \bibinfo{author}{{Budano}, A.}, \bibinfo{author}{{Camarri}, P.},
  \bibinfo{author}{{Cao}, Z.}, \bibinfo{author}{{Cardarelli}, R.},
  \bibinfo{author}{{Catalanotti}, S.}, \bibinfo{author}{{Chen}, S.~Z.},
  \bibinfo{author}{{Chen}, T.~L.}, \bibinfo{author}{{Creti}, P.},
  \bibinfo{author}{{Cui}, S.~W.}, \bibinfo{author}{{Dai}, B.~Z.},
  \bibinfo{author}{{D'Amone}, A.}, \bibinfo{author}{{Danzengluobu}},
  \bibinfo{author}{{De Mitri}, I.}, \bibinfo{author}{{D'Ettorre Piazzoli}, B.},
  \bibinfo{author}{{Di Girolamo}, T.}, \bibinfo{author}{{Di Sciascio}, G.},
  \bibinfo{author}{{Feng}, C.~F.}, \bibinfo{author}{{Feng}, Z.},
  \bibinfo{author}{{Feng}, Z.}, \bibinfo{author}{{Gou}, Q.~B.},
  \bibinfo{author}{{Guo}, Y.~Q.}, \bibinfo{author}{{He}, H.~H.},
  \bibinfo{author}{{Hu}, H.}, \bibinfo{author}{{Hu}, H.},
  \bibinfo{author}{{Iacovacci}, M.}, \bibinfo{author}{{Iuppa}, R.},
  \bibinfo{author}{{Jia}, H.~Y.}, \bibinfo{author}{{Labaciren}},
  \bibinfo{author}{{Li}, H.~J.}, \bibinfo{author}{{Liguori}, G.},
  \bibinfo{author}{{Liu}, C.}, \bibinfo{author}{{Liu}, J.},
  \bibinfo{author}{{Liu}, M.~Y.}, \bibinfo{author}{{Lu}, H.},
  \bibinfo{author}{{Ma}, L.~L.}, \bibinfo{author}{{Ma}, X.~H.},
  \bibinfo{author}{{Mancarella}, G.}, \bibinfo{author}{{Mari}, S.~M.},
  \bibinfo{author}{{Marsella}, G.}, \bibinfo{author}{{Martello}, D.},
  \bibinfo{author}{{Mastroianni}, S.}, \bibinfo{author}{{Montini}, P.},
  \bibinfo{author}{{Ning}, C.~C.}, \bibinfo{author}{{Panareo}, M.},
  \bibinfo{author}{{Perrone}, L.}, \bibinfo{author}{{Pistilli}, P.},
  \bibinfo{author}{{Ruggieri}, F.}, \bibinfo{author}{{Salvini}, P.},
  \bibinfo{author}{{Santonico}, R.}, \bibinfo{author}{{Shen}, P.~R.},
  \bibinfo{author}{{Sheng}, X.~D.}, \bibinfo{author}{{Shi}, F.},
  \bibinfo{author}{{Surdo}, A.}, \bibinfo{author}{{Tan}, Y.~H.},
  \bibinfo{author}{{Vallania}, P.}, \bibinfo{author}{{Vernetto}, S.},
  \bibinfo{author}{{Vigorito}, C.}, \bibinfo{author}{{Wang}, H.},
  \bibinfo{author}{{Wu}, C.~Y.}, \bibinfo{author}{{Wu}, H.~R.},
  \bibinfo{author}{{Xue}, L.}, \bibinfo{author}{{Yang}, Q.~Y.},
  \bibinfo{author}{{Yang}, X.~C.}, \bibinfo{author}{{Yao}, Z.~G.},
  \bibinfo{author}{{Yuan}, A.~F.}, \bibinfo{author}{{Zha}, M.},
  \bibinfo{author}{{Zhang}, H.~M.}, \bibinfo{author}{{Zhang}, L.},
  \bibinfo{author}{{Zhang}, X.~Y.}, \bibinfo{author}{{Zhang}, Y.},
  \bibinfo{author}{{Zhao}, J.}, \bibinfo{author}{{Zhaxiciren}},
  \bibinfo{author}{{Zhaxisangzhu}}, \bibinfo{author}{{Zhou}, X.~X.},
  \bibinfo{author}{{Zhu}, F.~R.}, \bibinfo{author}{{Zhu}, Q.~Q.},
  \bibinfo{author}{{Zizzi}, G.}, \& \bibinfo{author}{{ARGO-YBJ Collaboration}}
  (\bibinfo{year}{2014}).
\newblock \bibinfo{title}{{Identification of the TeV Gamma-Ray Source ARGO
  J2031+4157 with the Cygnus Cocoon}}.
\newblock {\it \bibinfo{journal}{The Astrophysical Journal}\/},  {\it
  \bibinfo{volume}{790}\/}\bibinfo{issue}{(2)}, \bibinfo{pages}{152}.
  \DOIprefix\doi{10.1088/0004-637X/790/2/152}.
  \href{http://arxiv.org/abs/1406.6436}{\tt arXiv:1406.6436}.
\bibitem[{{Binns} et~al.(2008){Binns}, {Wiedenbeck}, {Arnould}, {Cummings}, {de
  Nolfo}, {Goriely}, {Israel}, {Leske}, {Mewaldt}, {Stone} \& {von
  Rosenvinge}}]{2008NewAR..52..427B}
\bibinfo{author}{{Binns}, W.~R.}, \bibinfo{author}{{Wiedenbeck}, M.~E.},
  \bibinfo{author}{{Arnould}, M.}, \bibinfo{author}{{Cummings}, A.~C.},
  \bibinfo{author}{{de Nolfo}, G.~A.}, \bibinfo{author}{{Goriely}, S.},
  \bibinfo{author}{{Israel}, M.~H.}, \bibinfo{author}{{Leske}, R.~A.},
  \bibinfo{author}{{Mewaldt}, R.~A.}, \bibinfo{author}{{Stone}, E.~C.}, \&
  \bibinfo{author}{{von Rosenvinge}, T.~T.} (\bibinfo{year}{2008}).
\newblock \bibinfo{title}{{The OB association origin of galactic cosmic rays}}.
\newblock {\it \bibinfo{journal}{New Astronomy Reviews}\/},  {\it
  \bibinfo{volume}{52}\/}\bibinfo{issue}{(7-10)}, \bibinfo{pages}{427--430}.
  \DOIprefix\doi{10.1016/j.newar.2008.05.008}.
\bibitem[{{Bykov}(2001)}]{Bykov2001}
\bibinfo{author}{{Bykov}, A.~M.} (\bibinfo{year}{2001}).
\newblock \bibinfo{title}{{Particle Acceleration and Nonthermal Phenomena in
  Superbubbles}}.
\newblock {\it \bibinfo{journal}{Space Science Reviews}\/},  {\it
  \bibinfo{volume}{99}\/}, \bibinfo{pages}{317--326}.
\bibitem[{{Bykov}(2014)}]{Bykov2014}
\bibinfo{author}{{Bykov}, A.~M.} (\bibinfo{year}{2014}).
\newblock \bibinfo{title}{{Nonthermal particles and photons in starburst
  regions and superbubbles}}.
\newblock {\it \bibinfo{journal}{Astronomy \& Astrophysics Reviews}\/},  {\it
  \bibinfo{volume}{22}\/}, \bibinfo{pages}{77}.
  \DOIprefix\doi{10.1007/s00159-014-0077-8}.
\bibitem[{{Bykov} et~al.(2020){Bykov}, {Marcowith}, {Amato}, {Kalyashova},
  {Kruijssen} \& {Waxman}}]{2020SSRv..216...42B}
\bibinfo{author}{{Bykov}, A.~M.}, \bibinfo{author}{{Marcowith}, A.},
  \bibinfo{author}{{Amato}, E.}, \bibinfo{author}{{Kalyashova}, M.~E.},
  \bibinfo{author}{{Kruijssen}, J.~M.~D.}, \& \bibinfo{author}{{Waxman}, E.}
  (\bibinfo{year}{2020}).
\newblock \bibinfo{title}{{High-Energy Particles and Radiation in Star-Forming
  Regions}}.
\newblock {\it \bibinfo{journal}{Space Science Reviews}\/},  {\it
  \bibinfo{volume}{216}\/}\bibinfo{issue}{(3)}, \bibinfo{pages}{42}.
  \DOIprefix\doi{10.1007/s11214-020-00663-0}.
  \href{http://arxiv.org/abs/2003.11534}{\tt arXiv:2003.11534}.
\bibitem[{{Bykov} et~al.(2021){Bykov}, {Petrov}, {Kalyashova} \&
  {Troitsky}}]{PSR2032_21}
\bibinfo{author}{{Bykov}, A.~M.}, \bibinfo{author}{{Petrov}, A.~E.},
  \bibinfo{author}{{Kalyashova}, M.~E.}, \& \bibinfo{author}{{Troitsky}, S.~V.}
  (\bibinfo{year}{2021}).
\newblock \bibinfo{title}{{PeV Photon and Neutrino Flares from Galactic
  Gamma-Ray Binaries}}.
\newblock {\it \bibinfo{journal}{The Astrophysical Journal Letters}\/},  {\it
  \bibinfo{volume}{921}\/}\bibinfo{issue}{(1)}, \bibinfo{pages}{L10}.
  \DOIprefix\doi{10.3847/2041-8213/ac2f3d}.
  \href{http://arxiv.org/abs/2110.11189}{\tt arXiv:2110.11189}.
\bibitem[{{Bykov} \& {Toptygin}(1987)}]{BT87}
\bibinfo{author}{{Bykov}, A.~M.}, \& \bibinfo{author}{{Toptygin}, I.~N.}
  (\bibinfo{year}{1987}).
\newblock \bibinfo{title}{{Effect of Shocks on Interstellar Turbulence and
  Cosmic-Ray Dynamics}}.
\newblock {\it \bibinfo{journal}{Astrophysics and Space Science}\/},  {\it
  \bibinfo{volume}{138}\/}\bibinfo{issue}{(2)}, \bibinfo{pages}{341--354}.
  \DOIprefix\doi{10.1007/BF00637855}.
\bibitem[{{Bykov} \& {Toptygin}(1993)}]{bt93}
\bibinfo{author}{{Bykov}, A.~M.}, \& \bibinfo{author}{{Toptygin}, I.~N.}
  (\bibinfo{year}{1993}).
\newblock \bibinfo{title}{{Particle kinetics in highly turbulent plasmas
  (renormalization and self-consistent field methods)}}.
\newblock {\it \bibinfo{journal}{Physics Uspekhi}\/},  {\it
  \bibinfo{volume}{36}\/}, \bibinfo{pages}{1020--1052}.
\bibitem[{{Bykov} \& {Toptygin}(2001)}]{BT2001}
\bibinfo{author}{{Bykov}, A.~M.}, \& \bibinfo{author}{{Toptygin}, I.~N.}
  (\bibinfo{year}{2001}).
\newblock \bibinfo{title}{{A Model of Particle Acceleration to High Energies by
  Multiple Supernova Explosions in OB Associations}}.
\newblock {\it \bibinfo{journal}{Astronomy Letters}\/},  {\it
  \bibinfo{volume}{27}\/}\bibinfo{issue}{(10)}, \bibinfo{pages}{625--633}.
  \DOIprefix\doi{10.1134/1.1404456}.
\bibitem[{{Cao} et~al.(2021{\natexlab{a}}){Cao}, {Aharonian}, {An}, {Axikegu},
  {Bai}, {Bai}, {Bao}, {Bastieri}, {Bi}, {Bi}, {Cai}, {Cai}, {Cao}, {Chang},
  {Chang}, {Chen}, {Chen}, {Chen}, {Chen}, {Chen}, {Chen}, {Chen}, {Chen},
  {Chen}, {Chen}, {Chen}, {Chen}, {Chen}, {Chen}, {Cheng}, {Cheng}, {Cui},
  {Cui}, {Cui}, {D'Ettorre Piazzoli}, {Dai}, {Dai}, {Dai}, {Danzengluobu},
  {Volpe}, {Dong}, {Duan}, {Fan}, {Fan}, {Fan}, {Fang}, {Fang}, {Feng}, {Feng},
  {Feng}, {Feng}, {Gao}, {Gao}, {Gao}, {Gao}, {Gao}, {Ge}, {Geng}, {Gong},
  {Gou}, {Gu}, {Guo}, {Guo}, {Guo}, {Guo}, {Guo}, {Han}, {He}, {He}, {He},
  {He}, {He}, {He}, {Heller}, {Hor}, {Hou}, {Hu}, {Hu}, {Hu}, {Hu}, {Huang},
  {Huang}, {Huang}, {Huang}, {Huang}, {Huang}, {Ji}, {Ji}, {Jia}, {Jiang},
  {Jiang}, {Jin}, {Ke}, {Kuleshov}, {Levochkin}, {Li}, {Li}, {Li}, {Li}
  et~al.}]{LHAASO_J0341}
\bibinfo{author}{{Cao}, Z.}, \bibinfo{author}{{Aharonian}, F.},
  \bibinfo{author}{{An}, Q.}, \bibinfo{author}{{Axikegu}},
  \bibinfo{author}{{Bai}, L.~X.}, \bibinfo{author}{{Bai}, Y.~X.},
  \bibinfo{author}{{Bao}, Y.~W.}, \bibinfo{author}{{Bastieri}, D.},
  \bibinfo{author}{{Bi}, X.~J.}, \bibinfo{author}{{Bi}, Y.~J.},
  \bibinfo{author}{{Cai}, H.}, \bibinfo{author}{{Cai}, J.~T.},
  \bibinfo{author}{{Cao}, Z.}, \bibinfo{author}{{Chang}, J.},
  \bibinfo{author}{{Chang}, J.~F.}, \bibinfo{author}{{Chen}, B.~M.},
  \bibinfo{author}{{Chen}, E.~S.}, \bibinfo{author}{{Chen}, J.},
  \bibinfo{author}{{Chen}, L.}, \bibinfo{author}{{Chen}, L.},
  \bibinfo{author}{{Chen}, L.}, \bibinfo{author}{{Chen}, M.~J.},
  \bibinfo{author}{{Chen}, M.~L.}, \bibinfo{author}{{Chen}, Q.~H.},
  \bibinfo{author}{{Chen}, S.~H.}, \bibinfo{author}{{Chen}, S.~Z.},
  \bibinfo{author}{{Chen}, T.~L.}, \bibinfo{author}{{Chen}, X.~L.},
  \bibinfo{author}{{Chen}, Y.}, \bibinfo{author}{{Cheng}, N.},
  \bibinfo{author}{{Cheng}, Y.~D.}, \bibinfo{author}{{Cui}, S.~W.},
  \bibinfo{author}{{Cui}, X.~H.}, \bibinfo{author}{{Cui}, Y.~D.},
  \bibinfo{author}{{D'Ettorre Piazzoli}, B.}, \bibinfo{author}{{Dai}, B.~Z.},
  \bibinfo{author}{{Dai}, H.~L.}, \bibinfo{author}{{Dai}, Z.~G.},
  \bibinfo{author}{{Danzengluobu}}, \bibinfo{author}{{Volpe}, D.~d.},
  \bibinfo{author}{{Dong}, X.~J.}, \bibinfo{author}{{Duan}, K.~K.},
  \bibinfo{author}{{Fan}, J.~H.}, \bibinfo{author}{{Fan}, Y.~Z.},
  \bibinfo{author}{{Fan}, Z.~X.}, \bibinfo{author}{{Fang}, J.},
  \bibinfo{author}{{Fang}, K.}, \bibinfo{author}{{Feng}, C.~F.},
  \bibinfo{author}{{Feng}, L.}, \bibinfo{author}{{Feng}, S.~H.},
  \bibinfo{author}{{Feng}, Y.~L.}, \bibinfo{author}{{Gao}, B.},
  \bibinfo{author}{{Gao}, C.~D.}, \bibinfo{author}{{Gao}, L.~Q.},
  \bibinfo{author}{{Gao}, Q.}, \bibinfo{author}{{Gao}, W.},
  \bibinfo{author}{{Ge}, M.~M.}, \bibinfo{author}{{Geng}, L.~S.},
  \bibinfo{author}{{Gong}, G.~H.}, \bibinfo{author}{{Gou}, Q.~B.},
  \bibinfo{author}{{Gu}, M.~H.}, \bibinfo{author}{{Guo}, F.~L.},
  \bibinfo{author}{{Guo}, J.~G.}, \bibinfo{author}{{Guo}, X.~L.},
  \bibinfo{author}{{Guo}, Y.~Q.}, \bibinfo{author}{{Guo}, Y.~Y.},
  \bibinfo{author}{{Han}, Y.~A.}, \bibinfo{author}{{He}, H.~H.},
  \bibinfo{author}{{He}, H.~N.}, \bibinfo{author}{{He}, J.~C.},
  \bibinfo{author}{{He}, S.~L.}, \bibinfo{author}{{He}, X.~B.},
  \bibinfo{author}{{He}, Y.}, \bibinfo{author}{{Heller}, M.},
  \bibinfo{author}{{Hor}, Y.~K.}, \bibinfo{author}{{Hou}, C.},
  \bibinfo{author}{{Hu}, H.~B.}, \bibinfo{author}{{Hu}, S.},
  \bibinfo{author}{{Hu}, S.~C.}, \bibinfo{author}{{Hu}, X.~J.},
  \bibinfo{author}{{Huang}, D.~H.}, \bibinfo{author}{{Huang}, Q.~L.},
  \bibinfo{author}{{Huang}, W.~H.}, \bibinfo{author}{{Huang}, X.~T.},
  \bibinfo{author}{{Huang}, X.~Y.}, \bibinfo{author}{{Huang}, Z.~C.},
  \bibinfo{author}{{Ji}, F.}, \bibinfo{author}{{Ji}, X.~L.},
  \bibinfo{author}{{Jia}, H.~Y.}, \bibinfo{author}{{Jiang}, K.},
  \bibinfo{author}{{Jiang}, Z.~J.}, \bibinfo{author}{{Jin}, C.},
  \bibinfo{author}{{Ke}, T.}, \bibinfo{author}{{Kuleshov}, D.},
  \bibinfo{author}{{Levochkin}, K.}, \bibinfo{author}{{Li}, B.~B.},
  \bibinfo{author}{{Li}, C.}, \bibinfo{author}{{Li}, C.},
  \bibinfo{author}{{Li}, F.} et~al. (\bibinfo{year}{2021}{\natexlab{a}}).
\newblock \bibinfo{title}{{Discovery of a New Gamma-Ray Source, LHAASO
  J0341+5258, with Emission up to 200 TeV}}.
\newblock {\it \bibinfo{journal}{The Astrophysical Journal Letters}\/},  {\it
  \bibinfo{volume}{917}\/}\bibinfo{issue}{(1)}, \bibinfo{pages}{L4}.
  \DOIprefix\doi{10.3847/2041-8213/ac0fd5}.
\bibitem[{{Cao} et~al.(2021{\natexlab{b}}){Cao}, {Aharonian}, {An}, {Axikegu},
  {Bai}, {Bai}, {Bao}, {Bastieri}, {Bi}, {Bi}, {Cai}, {Cai}, {Cao}, {Chang},
  {Chang}, {Chen}, {Chen}, {Chen}, {Chen}, {Chen}, {Chen}, {Chen}, {Chen},
  {Chen}, {Chen}, {Chen}, {Chen}, {Chen}, {Cheng}, {Cheng}, {Cui}, {Cui},
  {Cui}, {Piazzoli}, {Dai}, {Dai}, {Dai}, {Dan-Zeng-Luo-Bu}, {Volpe}, {Dong},
  {Duan}, {Fan}, {Fan}, {Fan}, {Fang}, {Fang}, {Feng}, {Feng}, {Feng}, {Feng},
  {Gao}, {Gao}, {Gao}, {Gao}, {Gao}, {Ge}, {Geng}, {Gong}, {Gou}, {Gu}, {Guo},
  {Guo}, {Guo}, {Guo}, {Guo}, {Han}, {He}, {He}, {He}, {He}, {He}, {He},
  {Heller}, {Hor}, {Hou}, {Hu}, {Hu}, {Hu}, {Hu}, {Huang}, {Huang}, {Huang},
  {Huang}, {Huang}, {Huang}, {Ji}, {Ji}, {Jia}, {Jiang}, {Jiang}, {Jin}, {Ke},
  {Kuleshov}, {Levochkin}, {Li}, {Li}, {Li}, {Li}, {Li} et~al.}]{LHAASO_J2108}
\bibinfo{author}{{Cao}, Z.}, \bibinfo{author}{{Aharonian}, F.},
  \bibinfo{author}{{An}, Q.}, \bibinfo{author}{{Axikegu}},
  \bibinfo{author}{{Bai}, L.~X.}, \bibinfo{author}{{Bai}, Y.~X.},
  \bibinfo{author}{{Bao}, Y.~W.}, \bibinfo{author}{{Bastieri}, D.},
  \bibinfo{author}{{Bi}, X.~J.}, \bibinfo{author}{{Bi}, Y.~J.},
  \bibinfo{author}{{Cai}, H.}, \bibinfo{author}{{Cai}, J.~T.},
  \bibinfo{author}{{Cao}, Z.}, \bibinfo{author}{{Chang}, J.},
  \bibinfo{author}{{Chang}, J.~F.}, \bibinfo{author}{{Chen}, B.~M.},
  \bibinfo{author}{{Chen}, E.~S.}, \bibinfo{author}{{Chen}, J.},
  \bibinfo{author}{{Chen}, L.}, \bibinfo{author}{{Chen}, L.},
  \bibinfo{author}{{Chen}, M.~J.}, \bibinfo{author}{{Chen}, M.~L.},
  \bibinfo{author}{{Chen}, Q.~H.}, \bibinfo{author}{{Chen}, S.~H.},
  \bibinfo{author}{{Chen}, S.~Z.}, \bibinfo{author}{{Chen}, T.~L.},
  \bibinfo{author}{{Chen}, X.~L.}, \bibinfo{author}{{Chen}, Y.},
  \bibinfo{author}{{Cheng}, N.}, \bibinfo{author}{{Cheng}, Y.~D.},
  \bibinfo{author}{{Cui}, S.~W.}, \bibinfo{author}{{Cui}, X.~H.},
  \bibinfo{author}{{Cui}, Y.~D.}, \bibinfo{author}{{Piazzoli}, B.~D.},
  \bibinfo{author}{{Dai}, B.~Z.}, \bibinfo{author}{{Dai}, H.~L.},
  \bibinfo{author}{{Dai}, Z.~G.}, \bibinfo{author}{{Dan-Zeng-Luo-Bu}, D.-Z.},
  \bibinfo{author}{{Volpe}, D.~d.}, \bibinfo{author}{{Dong}, X.~J.},
  \bibinfo{author}{{Duan}, K.~K.}, \bibinfo{author}{{Fan}, J.~H.},
  \bibinfo{author}{{Fan}, Y.~Z.}, \bibinfo{author}{{Fan}, Z.~X.},
  \bibinfo{author}{{Fang}, J.}, \bibinfo{author}{{Fang}, K.},
  \bibinfo{author}{{Feng}, C.~F.}, \bibinfo{author}{{Feng}, L.},
  \bibinfo{author}{{Feng}, S.~H.}, \bibinfo{author}{{Feng}, Y.~L.},
  \bibinfo{author}{{Gao}, B.}, \bibinfo{author}{{Gao}, C.~D.},
  \bibinfo{author}{{Gao}, L.~Q.}, \bibinfo{author}{{Gao}, Q.},
  \bibinfo{author}{{Gao}, W.}, \bibinfo{author}{{Ge}, M.~M.},
  \bibinfo{author}{{Geng}, L.~S.}, \bibinfo{author}{{Gong}, G.~H.},
  \bibinfo{author}{{Gou}, Q.~B.}, \bibinfo{author}{{Gu}, M.~H.},
  \bibinfo{author}{{Guo}, F.~L.}, \bibinfo{author}{{Guo}, J.~G.},
  \bibinfo{author}{{Guo}, X.~L.}, \bibinfo{author}{{Guo}, Y.~Q.},
  \bibinfo{author}{{Guo}, Y.~Y.}, \bibinfo{author}{{Han}, Y.~A.},
  \bibinfo{author}{{He}, H.~H.}, \bibinfo{author}{{He}, H.~N.},
  \bibinfo{author}{{He}, J.~C.}, \bibinfo{author}{{He}, S.~L.},
  \bibinfo{author}{{He}, X.~B.}, \bibinfo{author}{{He}, Y.},
  \bibinfo{author}{{Heller}, M.}, \bibinfo{author}{{Hor}, Y.~K.},
  \bibinfo{author}{{Hou}, C.}, \bibinfo{author}{{Hu}, H.~B.},
  \bibinfo{author}{{Hu}, S.}, \bibinfo{author}{{Hu}, S.~C.},
  \bibinfo{author}{{Hu}, X.~J.}, \bibinfo{author}{{Huang}, D.~H.},
  \bibinfo{author}{{Huang}, Q.~L.}, \bibinfo{author}{{Huang}, W.~H.},
  \bibinfo{author}{{Huang}, X.~T.}, \bibinfo{author}{{Huang}, X.~Y.},
  \bibinfo{author}{{Huang}, Z.~C.}, \bibinfo{author}{{Ji}, F.},
  \bibinfo{author}{{Ji}, X.~L.}, \bibinfo{author}{{Jia}, H.~Y.},
  \bibinfo{author}{{Jiang}, K.}, \bibinfo{author}{{Jiang}, Z.~J.},
  \bibinfo{author}{{Jin}, C.}, \bibinfo{author}{{Ke}, T.},
  \bibinfo{author}{{Kuleshov}, D.}, \bibinfo{author}{{Levochkin}, K.},
  \bibinfo{author}{{Li}, B.~B.}, \bibinfo{author}{{Li}, C.},
  \bibinfo{author}{{Li}, C.}, \bibinfo{author}{{Li}, F.},
  \bibinfo{author}{{Li}, H.~B.} et~al. (\bibinfo{year}{2021}{\natexlab{b}}).
\newblock \bibinfo{title}{{Discovery of the Ultrahigh-energy Gamma-Ray Source
  LHAASO J2108+5157}}.
\newblock {\it \bibinfo{journal}{The Astrophysical Journal Letters}\/},  {\it
  \bibinfo{volume}{919}\/}\bibinfo{issue}{(2)}, \bibinfo{pages}{L22}.
  \DOIprefix\doi{10.3847/2041-8213/ac2579}.
\bibitem[{{Cao} et~al.(2021{\natexlab{c}}){Cao}, {Aharonian}, {An}, {Axikegu},
  {Bai}, {Bao}, {Bastieri}, {Bi}, {Bi}, {Cai}, {Cai}, {Cao}, {Chang}, {Chang},
  {Chang}, {Chen}, {Chen}, {Chen}, {Chen}, {Chen}, {Chen}, {Chen}, {Chen},
  {Chen}, {Chen}, {Chen}, {Chen}, {Chen}, {Cheng}, {Cheng}, {Cui}, {Cui},
  {Cui}, {Dai}, {Dai}, {Dai}, {Danzengluobu}, {della Volpe}, {D'Ettorre
  Piazzoli}, {Dong}, {Fan}, {Fan}, {Fan}, {Fang}, {Fang}, {Feng}, {Feng},
  {Feng}, {Feng}, {Gao}, {Gao}, {Gao}, {Gao}, {Ge}, {Geng}, {Gong}, {Gou},
  {Gu}, {Guo}, {Guo}, {Guo}, {Guo}, {Han}, {He}, {He}, {He}, {He}, {He}, {He},
  {Heller}, {Hor}, {Hou}, {Hou}, {Hu}, {Hu}, {Hu}, {Hu}, {Huang}, {Huang},
  {Huang}, {Huang}, {Huang}, {Ji}, {Ji}, {Jia}, {Jiang}, {Jiang}, {Jin},
  {Kuleshov}, {Levochkin}, {Li}, {Li}, {Li}, {Li}, {Li}, {Li}, {Li}, {Li}, {Li}
  et~al.}]{LHAASO2021}
\bibinfo{author}{{Cao}, Z.}, \bibinfo{author}{{Aharonian}, F.~A.},
  \bibinfo{author}{{An}, Q.}, \bibinfo{author}{{Axikegu}, L.~X., Bai},
  \bibinfo{author}{{Bai}, Y.~X.}, \bibinfo{author}{{Bao}, Y.~W.},
  \bibinfo{author}{{Bastieri}, D.}, \bibinfo{author}{{Bi}, X.~J.},
  \bibinfo{author}{{Bi}, Y.~J.}, \bibinfo{author}{{Cai}, H.},
  \bibinfo{author}{{Cai}, J.~T.}, \bibinfo{author}{{Cao}, Z.},
  \bibinfo{author}{{Chang}, J.}, \bibinfo{author}{{Chang}, J.~F.},
  \bibinfo{author}{{Chang}, X.~C.}, \bibinfo{author}{{Chen}, B.~M.},
  \bibinfo{author}{{Chen}, J.}, \bibinfo{author}{{Chen}, L.},
  \bibinfo{author}{{Chen}, L.}, \bibinfo{author}{{Chen}, L.},
  \bibinfo{author}{{Chen}, M.~J.}, \bibinfo{author}{{Chen}, M.~L.},
  \bibinfo{author}{{Chen}, Q.~H.}, \bibinfo{author}{{Chen}, S.~H.},
  \bibinfo{author}{{Chen}, S.~Z.}, \bibinfo{author}{{Chen}, T.~L.},
  \bibinfo{author}{{Chen}, X.~L.}, \bibinfo{author}{{Chen}, Y.},
  \bibinfo{author}{{Cheng}, N.}, \bibinfo{author}{{Cheng}, Y.~D.},
  \bibinfo{author}{{Cui}, S.~W.}, \bibinfo{author}{{Cui}, X.~H.},
  \bibinfo{author}{{Cui}, Y.~D.}, \bibinfo{author}{{Dai}, B.~Z.},
  \bibinfo{author}{{Dai}, H.~L.}, \bibinfo{author}{{Dai}, Z.~G.},
  \bibinfo{author}{{Danzengluobu}}, \bibinfo{author}{{della Volpe}, D.},
  \bibinfo{author}{{D'Ettorre Piazzoli}, B.}, \bibinfo{author}{{Dong}, X.~J.},
  \bibinfo{author}{{Fan}, J.~H.}, \bibinfo{author}{{Fan}, Y.~Z.},
  \bibinfo{author}{{Fan}, Z.~X.}, \bibinfo{author}{{Fang}, J.},
  \bibinfo{author}{{Fang}, K.}, \bibinfo{author}{{Feng}, C.~F.},
  \bibinfo{author}{{Feng}, L.}, \bibinfo{author}{{Feng}, S.~H.},
  \bibinfo{author}{{Feng}, Y.~L.}, \bibinfo{author}{{Gao}, B.},
  \bibinfo{author}{{Gao}, C.~D.}, \bibinfo{author}{{Gao}, Q.},
  \bibinfo{author}{{Gao}, W.}, \bibinfo{author}{{Ge}, M.~M.},
  \bibinfo{author}{{Geng}, L.~S.}, \bibinfo{author}{{Gong}, G.~H.},
  \bibinfo{author}{{Gou}, Q.~B.}, \bibinfo{author}{{Gu}, M.~H.},
  \bibinfo{author}{{Guo}, J.~G.}, \bibinfo{author}{{Guo}, X.~L.},
  \bibinfo{author}{{Guo}, Y.~Q.}, \bibinfo{author}{{Guo}, Y.~Y.},
  \bibinfo{author}{{Han}, Y.~A.}, \bibinfo{author}{{He}, H.~H.},
  \bibinfo{author}{{He}, H.~N.}, \bibinfo{author}{{He}, J.~C.},
  \bibinfo{author}{{He}, S.~L.}, \bibinfo{author}{{He}, X.~B.},
  \bibinfo{author}{{He}, Y.}, \bibinfo{author}{{Heller}, M.},
  \bibinfo{author}{{Hor}, Y.~K.}, \bibinfo{author}{{Hou}, C.},
  \bibinfo{author}{{Hou}, X.}, \bibinfo{author}{{Hu}, H.~B.},
  \bibinfo{author}{{Hu}, S.}, \bibinfo{author}{{Hu}, S.~C.},
  \bibinfo{author}{{Hu}, X.~J.}, \bibinfo{author}{{Huang}, D.~H.},
  \bibinfo{author}{{Huang}, Q.~L.}, \bibinfo{author}{{Huang}, W.~H.},
  \bibinfo{author}{{Huang}, X.~T.}, \bibinfo{author}{{Huang}, Z.~C.},
  \bibinfo{author}{{Ji}, F.}, \bibinfo{author}{{Ji}, X.~L.},
  \bibinfo{author}{{Jia}, H.~Y.}, \bibinfo{author}{{Jiang}, K.},
  \bibinfo{author}{{Jiang}, Z.~J.}, \bibinfo{author}{{Jin}, C.},
  \bibinfo{author}{{Kuleshov}, D.}, \bibinfo{author}{{Levochkin}, K.},
  \bibinfo{author}{{Li}, B.~B.}, \bibinfo{author}{{Li}, C.},
  \bibinfo{author}{{Li}, C.}, \bibinfo{author}{{Li}, F.},
  \bibinfo{author}{{Li}, H.~B.}, \bibinfo{author}{{Li}, H.~C.},
  \bibinfo{author}{{Li}, H.~Y.}, \bibinfo{author}{{Li}, J.},
  \bibinfo{author}{{Li}, K.} et~al. (\bibinfo{year}{2021}{\natexlab{c}}).
\newblock \bibinfo{title}{{Ultrahigh-energy photons up to 1.4 petaelectronvolts
  from 12 {\ensuremath{\gamma}}-ray Galactic sources}}.
\newblock {\it \bibinfo{journal}{\nat}\/},  {\it
  \bibinfo{volume}{594}\/}\bibinfo{issue}{(7861)}, \bibinfo{pages}{33--36}.
  \DOIprefix\doi{10.1038/s41586-021-03498-z}.
\bibitem[{{Cesarsky} \& {Montmerle}(1983)}]{1983SSRv...36..173C}
\bibinfo{author}{{Cesarsky}, C.~J.}, \& \bibinfo{author}{{Montmerle}, T.}
  (\bibinfo{year}{1983}).
\newblock \bibinfo{title}{{Gamma-Rays from Active Regions in the Galaxy - the
  Possible Contribution of Stellar Winds}}.
\newblock {\it \bibinfo{journal}{Space Science Reviews}\/},  {\it
  \bibinfo{volume}{36}\/}\bibinfo{issue}{(2)}, \bibinfo{pages}{173--193}.
  \DOIprefix\doi{10.1007/BF00167503}.
\bibitem[{{Churazov} et~al.(2021){Churazov}, {Khabibullin}, {Bykov}, {Chugai},
  {Sunyaev} \& {Zinchenko}}]{G116}
\bibinfo{author}{{Churazov}, E.~M.}, \bibinfo{author}{{Khabibullin}, I.~I.},
  \bibinfo{author}{{Bykov}, A.~M.}, \bibinfo{author}{{Chugai}, N.~N.},
  \bibinfo{author}{{Sunyaev}, R.~A.}, \& \bibinfo{author}{{Zinchenko}, I.~I.}
  (\bibinfo{year}{2021}).
\newblock \bibinfo{title}{{SRG/eROSITA discovery of a large circular SNR
  candidate G116.6-26.1: SN Ia explosion probing the gas of the Milky Way
  halo?}}
\newblock {\it \bibinfo{journal}{Monthly Notices of the Royal Astronomical
  Society}\/},  {\it \bibinfo{volume}{507}\/}\bibinfo{issue}{(1)},
  \bibinfo{pages}{971--982}. \DOIprefix\doi{10.1093/mnras/stab2125}.
  \href{http://arxiv.org/abs/2106.09454}{\tt arXiv:2106.09454}.
\bibitem[{{Dzhappuev} et~al.(2021){Dzhappuev}, {Afashokov}, {Dzaparova},
  {Dzhatdoev}, {Gorbacheva}, {Karpikov}, {Khadzhiev}, {Klimenko}, {Kudzhaev},
  {Kurenya}, {Lidvansky}, {Mikhailova}, {Petkov}, {Podlesnyi}, {Romanenko},
  {Rubtsov}, {Troitsky}, {Unatlokov}, {Vaiman}, {Yanin}, {Zhezher},
  {Zhuravleva} \& {Carpet-3 Group}}]{Dzhappuev2021}
\bibinfo{author}{{Dzhappuev}, D.~D.}, \bibinfo{author}{{Afashokov}, Y.~Z.},
  \bibinfo{author}{{Dzaparova}, I.~M.}, \bibinfo{author}{{Dzhatdoev}, T.~A.},
  \bibinfo{author}{{Gorbacheva}, E.~A.}, \bibinfo{author}{{Karpikov}, I.~S.},
  \bibinfo{author}{{Khadzhiev}, M.~M.}, \bibinfo{author}{{Klimenko}, N.~F.},
  \bibinfo{author}{{Kudzhaev}, A.~U.}, \bibinfo{author}{{Kurenya}, A.~N.},
  \bibinfo{author}{{Lidvansky}, A.~S.}, \bibinfo{author}{{Mikhailova}, O.~I.},
  \bibinfo{author}{{Petkov}, V.~B.}, \bibinfo{author}{{Podlesnyi}, E.~I.},
  \bibinfo{author}{{Romanenko}, V.~S.}, \bibinfo{author}{{Rubtsov}, G.~I.},
  \bibinfo{author}{{Troitsky}, S.~V.}, \bibinfo{author}{{Unatlokov}, I.~B.},
  \bibinfo{author}{{Vaiman}, I.~A.}, \bibinfo{author}{{Yanin}, A.~F.},
  \bibinfo{author}{{Zhezher}, Y.~V.}, \bibinfo{author}{{Zhuravleva}, K.~V.}, \&
  \bibinfo{author}{{Carpet-3 Group}} (\bibinfo{year}{2021}).
\newblock \bibinfo{title}{{Observation of Photons above 300 TeV Associated with
  a High-energy Neutrino from the Cygnus Region}}.
\newblock {\it \bibinfo{journal}{The Astrophysical Journal Letters}\/},  {\it
  \bibinfo{volume}{916}\/}\bibinfo{issue}{(2)}, \bibinfo{pages}{L22}.
  \DOIprefix\doi{10.3847/2041-8213/ac14b2}.
  \href{http://arxiv.org/abs/2105.07242}{\tt arXiv:2105.07242}.
\bibitem[{{Ferrand} \& {Marcowith}(2010)}]{2010A&A...510A.101F}
\bibinfo{author}{{Ferrand}, G.}, \& \bibinfo{author}{{Marcowith}, A.}
  (\bibinfo{year}{2010}).
\newblock \bibinfo{title}{{On the shape of the spectrum of cosmic rays
  accelerated inside superbubbles}}.
\newblock {\it \bibinfo{journal}{Astronomy \& Astrophysics}\/},  {\it
  \bibinfo{volume}{510}\/}, \bibinfo{pages}{A101}.
  \DOIprefix\doi{10.1051/0004-6361/200913520}.
  \href{http://arxiv.org/abs/0911.4457}{\tt arXiv:0911.4457}.
\bibitem[{{Foreman-Mackey} et~al.(2013){Foreman-Mackey}, {Hogg}, {Lang} \&
  {Goodman}}]{MCMC}
\bibinfo{author}{{Foreman-Mackey}, D.}, \bibinfo{author}{{Hogg}, D.~W.},
  \bibinfo{author}{{Lang}, D.}, \& \bibinfo{author}{{Goodman}, J.}
  (\bibinfo{year}{2013}).
\newblock \bibinfo{title}{{emcee: The MCMC Hammer}}.
\newblock {\it \bibinfo{journal}{Publications of the Astronomical Society of
  the Pacific}\/},  {\it \bibinfo{volume}{125}\/}\bibinfo{issue}{(925)},
  \bibinfo{pages}{306}. \DOIprefix\doi{10.1086/670067}.
  \href{http://arxiv.org/abs/1202.3665}{\tt arXiv:1202.3665}.
\bibitem[{{Gupta} et~al.(2020){Gupta}, {Nath}, {Sharma} \&
  {Eichler}}]{2020MNRAS.493.3159G}
\bibinfo{author}{{Gupta}, S.}, \bibinfo{author}{{Nath}, B.~B.},
  \bibinfo{author}{{Sharma}, P.}, \& \bibinfo{author}{{Eichler}, D.}
  (\bibinfo{year}{2020}).
\newblock \bibinfo{title}{{Realistic modelling of wind and supernovae shocks in
  star clusters: addressing $^{22}$Ne/$^{20}$Ne and other problems in Galactic
  cosmic rays}}.
\newblock {\it \bibinfo{journal}{Monthly Notices of the Royal Astronomical
  Society}\/},  {\it \bibinfo{volume}{493}\/}\bibinfo{issue}{(3)},
  \bibinfo{pages}{3159--3177}. \DOIprefix\doi{10.1093/mnras/staa286}.
  \href{http://arxiv.org/abs/1910.10168}{\tt arXiv:1910.10168}.
\bibitem[{{H.~E.~S.~S. Collaboration} et~al.(2011){H.~E.~S.~S. Collaboration},
  {Abramowski}, {Acero}, {Aharonian}, {Akhperjanian}, {Anton}, {Barnacka},
  {Barres de Almeida}, {Bazer-Bachi}, {Becherini}, {Becker}, {Behera},
  {Bernl{\"o}hr}, {Bochow}, {Boisson}, {Bolmont}, {Bordas}, {Borrel},
  {Brucker}, {Brun}, {Brun}, {Bulik}, {B{\"u}sching}, {Boutelier}, {Casanova},
  {Cerruti}, {Chadwick}, {Charbonnier}, {Chaves}, {Cheesebrough}, {Conrad},
  {Chounet}, {Clapson}, {Coignet}, {Dalton}, {Daniel}, {Davids}, {Degrange},
  {Deil}, {Dickinson}, {Djannati-Ata{\"\i}}, {Domainko}, {Drury}, {Dubois},
  {Dubus}, {Dyks}, {Dyrda}, {Egberts}, {Eger}, {Espigat}, {Fallon}, {Farnier},
  {Fegan}, {Feinstein}, {Fernandes}, {Fiasson}, {F{\"o}rster}, {Fontaine},
  {F{\"u}{\ss}ling}, {Gabici}, {Gallant}, {G{\'e}rard}, {Gerbig}, {Giebels},
  {Glicenstein}, {Gl{\"u}ck}, {Goret}, {G{\"o}ring}, {Hague}, {Hampf},
  {Hauser}, {Heinz}, {Heinzelmann}, {Henri}, {Hermann}, {Hinton}, {Hoffmann},
  {Hofmann}, {Hofverberg}, {Holleran}, {Hoppe}, {Horns}, {Jacholkowska}, {de
  Jager}, {Jahn}, {Jung}, {Katarzy{\'n}ski}, {Katz}, {Kaufmann}, {Kerschhaggl},
  {Khangulyan}, {Kh{\'e}lifi}, {Keogh}, {Klochkov}, {Klu{\'z}niak}, {Kneiske},
  {Komin}, {Kosack}, {Kossakowski} et~al.}]{HessWd2}
\bibinfo{author}{{H.~E.~S.~S. Collaboration}}, \bibinfo{author}{{Abramowski},
  A.}, \bibinfo{author}{{Acero}, F.}, \bibinfo{author}{{Aharonian}, F.},
  \bibinfo{author}{{Akhperjanian}, A.~G.}, \bibinfo{author}{{Anton}, G.},
  \bibinfo{author}{{Barnacka}, A.}, \bibinfo{author}{{Barres de Almeida}, U.},
  \bibinfo{author}{{Bazer-Bachi}, A.~R.}, \bibinfo{author}{{Becherini}, Y.},
  \bibinfo{author}{{Becker}, J.}, \bibinfo{author}{{Behera}, B.},
  \bibinfo{author}{{Bernl{\"o}hr}, K.}, \bibinfo{author}{{Bochow}, A.},
  \bibinfo{author}{{Boisson}, C.}, \bibinfo{author}{{Bolmont}, J.},
  \bibinfo{author}{{Bordas}, P.}, \bibinfo{author}{{Borrel}, V.},
  \bibinfo{author}{{Brucker}, J.}, \bibinfo{author}{{Brun}, F.},
  \bibinfo{author}{{Brun}, P.}, \bibinfo{author}{{Bulik}, T.},
  \bibinfo{author}{{B{\"u}sching}, I.}, \bibinfo{author}{{Boutelier}, T.},
  \bibinfo{author}{{Casanova}, S.}, \bibinfo{author}{{Cerruti}, M.},
  \bibinfo{author}{{Chadwick}, P.~M.}, \bibinfo{author}{{Charbonnier}, A.},
  \bibinfo{author}{{Chaves}, R.~C.~G.}, \bibinfo{author}{{Cheesebrough}, A.},
  \bibinfo{author}{{Conrad}, J.}, \bibinfo{author}{{Chounet}, L.~M.},
  \bibinfo{author}{{Clapson}, A.~C.}, \bibinfo{author}{{Coignet}, G.},
  \bibinfo{author}{{Dalton}, M.}, \bibinfo{author}{{Daniel}, M.~K.},
  \bibinfo{author}{{Davids}, I.~D.}, \bibinfo{author}{{Degrange}, B.},
  \bibinfo{author}{{Deil}, C.}, \bibinfo{author}{{Dickinson}, H.~J.},
  \bibinfo{author}{{Djannati-Ata{\"\i}}, A.}, \bibinfo{author}{{Domainko}, W.},
  \bibinfo{author}{{Drury}, L.~O.}, \bibinfo{author}{{Dubois}, F.},
  \bibinfo{author}{{Dubus}, G.}, \bibinfo{author}{{Dyks}, J.},
  \bibinfo{author}{{Dyrda}, M.}, \bibinfo{author}{{Egberts}, K.},
  \bibinfo{author}{{Eger}, P.}, \bibinfo{author}{{Espigat}, P.},
  \bibinfo{author}{{Fallon}, L.}, \bibinfo{author}{{Farnier}, C.},
  \bibinfo{author}{{Fegan}, S.}, \bibinfo{author}{{Feinstein}, F.},
  \bibinfo{author}{{Fernandes}, M.~V.}, \bibinfo{author}{{Fiasson}, A.},
  \bibinfo{author}{{F{\"o}rster}, A.}, \bibinfo{author}{{Fontaine}, G.},
  \bibinfo{author}{{F{\"u}{\ss}ling}, M.}, \bibinfo{author}{{Gabici}, S.},
  \bibinfo{author}{{Gallant}, Y.~A.}, \bibinfo{author}{{G{\'e}rard}, L.},
  \bibinfo{author}{{Gerbig}, D.}, \bibinfo{author}{{Giebels}, B.},
  \bibinfo{author}{{Glicenstein}, J.~F.}, \bibinfo{author}{{Gl{\"u}ck}, B.},
  \bibinfo{author}{{Goret}, P.}, \bibinfo{author}{{G{\"o}ring}, D.},
  \bibinfo{author}{{Hague}, J.~D.}, \bibinfo{author}{{Hampf}, D.},
  \bibinfo{author}{{Hauser}, M.}, \bibinfo{author}{{Heinz}, S.},
  \bibinfo{author}{{Heinzelmann}, G.}, \bibinfo{author}{{Henri}, G.},
  \bibinfo{author}{{Hermann}, G.}, \bibinfo{author}{{Hinton}, J.~A.},
  \bibinfo{author}{{Hoffmann}, A.}, \bibinfo{author}{{Hofmann}, W.},
  \bibinfo{author}{{Hofverberg}, P.}, \bibinfo{author}{{Holleran}, M.},
  \bibinfo{author}{{Hoppe}, S.}, \bibinfo{author}{{Horns}, D.},
  \bibinfo{author}{{Jacholkowska}, A.}, \bibinfo{author}{{de Jager}, O.~C.},
  \bibinfo{author}{{Jahn}, C.}, \bibinfo{author}{{Jung}, I.},
  \bibinfo{author}{{Katarzy{\'n}ski}, K.}, \bibinfo{author}{{Katz}, U.},
  \bibinfo{author}{{Kaufmann}, S.}, \bibinfo{author}{{Kerschhaggl}, M.},
  \bibinfo{author}{{Khangulyan}, D.}, \bibinfo{author}{{Kh{\'e}lifi}, B.},
  \bibinfo{author}{{Keogh}, D.}, \bibinfo{author}{{Klochkov}, D.},
  \bibinfo{author}{{Klu{\'z}niak}, W.}, \bibinfo{author}{{Kneiske}, T.},
  \bibinfo{author}{{Komin}, N.}, \bibinfo{author}{{Kosack}, K.},
  \bibinfo{author}{{Kossakowski}, R.} et~al. (\bibinfo{year}{2011}).
\newblock \bibinfo{title}{{Revisiting the Westerlund 2 field with the HESS
  telescope array}}.
\newblock {\it \bibinfo{journal}{Astronomy \& Astrophysics}\/},  {\it
  \bibinfo{volume}{525}\/}, \bibinfo{pages}{A46}.
  \DOIprefix\doi{10.1051/0004-6361/201015290}.
  \href{http://arxiv.org/abs/1009.3012}{\tt arXiv:1009.3012}.
\bibitem[{{Hanson}(2003)}]{Hanson2003}
\bibinfo{author}{{Hanson}, M.~M.} (\bibinfo{year}{2003}).
\newblock \bibinfo{title}{{A Study of Cygnus OB2: Pointing the Way toward
  Finding Our Galaxy's Super-Star Clusters}}.
\newblock {\it \bibinfo{journal}{The Astrophysical Journal}\/},  {\it
  \bibinfo{volume}{597}\/}\bibinfo{issue}{(2)}, \bibinfo{pages}{957--969}.
  \DOIprefix\doi{10.1086/378508}.
  \href{http://arxiv.org/abs/astro-ph/0307540}{\tt arXiv:astro-ph/0307540}.
\bibitem[{Hona et~al.(2019)Hona, Fleischhack \& Huentemeyer}]{Hona2019}
\bibinfo{author}{Hona, B.}, \bibinfo{author}{Fleischhack, H.}, \&
  \bibinfo{author}{Huentemeyer, P.} (\bibinfo{year}{2019}).
\newblock \bibinfo{title}{{Testing the Limits of Particle Acceleration in
  Cygnus OB2 with HAWC}}.
\newblock In {\it \bibinfo{booktitle}{Proceedings of 36th International Cosmic
  Ray Conference {\textemdash} PoS(ICRC2019)}\/} (p. \bibinfo{pages}{699}).
\newblock volume \bibinfo{volume}{358}.
\newblock \DOIprefix\doi{10.22323/1.358.0699}.
\bibitem[{{Joubaud} et~al.(2020){Joubaud}, {Grenier}, {Casandjian}, {Tolksdorf}
  \& {Schlickeiser}}]{2020A&A...635A..96J}
\bibinfo{author}{{Joubaud}, T.}, \bibinfo{author}{{Grenier}, I.~A.},
  \bibinfo{author}{{Casandjian}, J.~M.}, \bibinfo{author}{{Tolksdorf}, T.}, \&
  \bibinfo{author}{{Schlickeiser}, R.} (\bibinfo{year}{2020}).
\newblock \bibinfo{title}{{The cosmic-ray content of the Orion-Eridanus
  superbubble}}.
\newblock {\it \bibinfo{journal}{Astronomy \& Astrophysics}\/},  {\it
  \bibinfo{volume}{635}\/}, \bibinfo{pages}{A96}.
  \DOIprefix\doi{10.1051/0004-6361/201937205}.
  \href{http://arxiv.org/abs/2001.10139}{\tt arXiv:2001.10139}.
\bibitem[{{Kafexhiu} et~al.(2014){Kafexhiu}, {Aharonian}, {Taylor} \&
  {Vila}}]{KafexhiuATV2014}
\bibinfo{author}{{Kafexhiu}, E.}, \bibinfo{author}{{Aharonian}, F.},
  \bibinfo{author}{{Taylor}, A.~M.}, \& \bibinfo{author}{{Vila}, G.~S.}
  (\bibinfo{year}{2014}).
\newblock \bibinfo{title}{{Parametrization of gamma-ray production cross
  sections for p p interactions in a broad proton energy range from the
  kinematic threshold to PeV energies}}.
\newblock {\it \bibinfo{journal}{Physical Review D}\/},  {\it
  \bibinfo{volume}{90}\/}\bibinfo{issue}{(12)}, \bibinfo{pages}{123014}.
  \DOIprefix\doi{10.1103/PhysRevD.90.123014}.
  \href{http://arxiv.org/abs/1406.7369}{\tt arXiv:1406.7369}.
\bibitem[{{Kalyashova} et~al.(2019){Kalyashova}, {Bykov}, {Osipov}, {Ellison}
  \& {Badmaev}}]{2019JPhCS1400b2011K}
\bibinfo{author}{{Kalyashova}, M.~E.}, \bibinfo{author}{{Bykov}, A.~M.},
  \bibinfo{author}{{Osipov}, S.~M.}, \bibinfo{author}{{Ellison}, D.~C.}, \&
  \bibinfo{author}{{Badmaev}, D.~V.} (\bibinfo{year}{2019}).
\newblock \bibinfo{title}{{Wolf-Rayet stars in young massive star clusters as
  potential sources of Galactic cosmic rays}}.
\newblock In {\it \bibinfo{booktitle}{Journal of Physics Conference Series}\/}
  (p. \bibinfo{pages}{022011}).
\newblock volume \bibinfo{volume}{1400}.
\newblock \DOIprefix\doi{10.1088/1742-6596/1400/2/022011}.
  \href{http://arxiv.org/abs/1910.08602}{\tt arXiv:1910.08602}.
\bibitem[{{Kelner} et~al.(2006){Kelner}, {Aharonian} \& {Bugayov}}]{Kelner2006}
\bibinfo{author}{{Kelner}, S.~R.}, \bibinfo{author}{{Aharonian}, F.~A.}, \&
  \bibinfo{author}{{Bugayov}, V.~V.} (\bibinfo{year}{2006}).
\newblock \bibinfo{title}{{Energy spectra of gamma rays, electrons, and
  neutrinos produced at proton-proton interactions in the very high energy
  regime}}.
\newblock {\it \bibinfo{journal}{Physical Review D}\/},  {\it
  \bibinfo{volume}{74}\/}\bibinfo{issue}{(3)}, \bibinfo{pages}{034018}.
  \DOIprefix\doi{10.1103/PhysRevD.74.034018}.
  \href{http://arxiv.org/abs/arXiv:astro-ph/0606058}{\tt
  arXiv:arXiv:astro-ph/0606058}.
\bibitem[{{Kn{\"o}dlseder}(2000)}]{2000A&A...360..539K}
\bibinfo{author}{{Kn{\"o}dlseder}, J.} (\bibinfo{year}{2000}).
\newblock \bibinfo{title}{{Cygnus OB2 - a young globular cluster in the Milky
  Way}}.
\newblock {\it \bibinfo{journal}{Astronomy \& Astrophysics}\/},  {\it
  \bibinfo{volume}{360}\/}, \bibinfo{pages}{539--548}.
  \href{http://arxiv.org/abs/astro-ph/0007442}{\tt arXiv:astro-ph/0007442}.
\bibitem[{{Lemoine}(2019)}]{2019PhRvD..99h3006L}
\bibinfo{author}{{Lemoine}, M.} (\bibinfo{year}{2019}).
\newblock \bibinfo{title}{{Generalized Fermi acceleration}}.
\newblock {\it \bibinfo{journal}{Physical Review D}\/},  {\it
  \bibinfo{volume}{99}\/}\bibinfo{issue}{(8)}, \bibinfo{pages}{083006}.
  \DOIprefix\doi{10.1103/PhysRevD.99.083006}.
  \href{http://arxiv.org/abs/1903.05917}{\tt arXiv:1903.05917}.
\bibitem[{{Lemoine} \& {Waxman}(2009)}]{Lemoine2009}
\bibinfo{author}{{Lemoine}, M.}, \& \bibinfo{author}{{Waxman}, E.}
  (\bibinfo{year}{2009}).
\newblock \bibinfo{title}{{Anisotropy vs chemical composition at ultra-high
  energies}}.
\newblock {\it \bibinfo{journal}{Journal of Cosmology and Astroparticle
  Physics}\/},  {\it \bibinfo{volume}{2009}\/}\bibinfo{issue}{(11)},
  \bibinfo{pages}{009}. \DOIprefix\doi{10.1088/1475-7516/2009/11/009}.
  \href{http://arxiv.org/abs/0907.1354}{\tt arXiv:0907.1354}.
\bibitem[{{Lingenfelter}(2018)}]{Lingenfelter18}
\bibinfo{author}{{Lingenfelter}, R.~E.} (\bibinfo{year}{2018}).
\newblock \bibinfo{title}{{Cosmic rays from supernova remnants and
  superbubbles}}.
\newblock {\it \bibinfo{journal}{Advances in Space Research}\/},  {\it
  \bibinfo{volume}{62}\/}\bibinfo{issue}{(10)}, \bibinfo{pages}{2750--2763}.
  \DOIprefix\doi{10.1016/j.asr.2017.04.006}.
  \href{http://arxiv.org/abs/1807.09726}{\tt arXiv:1807.09726}.
\bibitem[{{Liu} \& {Wang}(2021)}]{Liu2021}
\bibinfo{author}{{Liu}, R.-Y.}, \& \bibinfo{author}{{Wang}, X.-Y.}
  (\bibinfo{year}{2021}).
\newblock \bibinfo{title}{{Origin of Galactic Sub-PeV Diffuse Gamma-Ray
  Emission: Constraints from High-energy Neutrino Observations}}.
\newblock {\it \bibinfo{journal}{The Astrophysical Journal Letters}\/},  {\it
  \bibinfo{volume}{914}\/}\bibinfo{issue}{(1)}, \bibinfo{pages}{L7}.
  \DOIprefix\doi{10.3847/2041-8213/ac02c5}.
  \href{http://arxiv.org/abs/2104.05609}{\tt arXiv:2104.05609}.
\bibitem[{{Malkov}(2017)}]{2017PhRvD..95b3007M}
\bibinfo{author}{{Malkov}, M.~A.} (\bibinfo{year}{2017}).
\newblock \bibinfo{title}{{Exact solution of the Fokker-Planck equation for
  isotropic scattering}}.
\newblock {\it \bibinfo{journal}{Physical Review D}\/},  {\it
  \bibinfo{volume}{95}\/}\bibinfo{issue}{(2)}, \bibinfo{pages}{023007}.
  \DOIprefix\doi{10.1103/PhysRevD.95.023007}.
  \href{http://arxiv.org/abs/1610.01584}{\tt arXiv:1610.01584}.
\bibitem[{{Malkov} et~al.(2013){Malkov}, {Diamond}, {Sagdeev}, {Aharonian} \&
  {Moskalenko}}]{2013ApJ...768...73M}
\bibinfo{author}{{Malkov}, M.~A.}, \bibinfo{author}{{Diamond}, P.~H.},
  \bibinfo{author}{{Sagdeev}, R.~Z.}, \bibinfo{author}{{Aharonian}, F.~A.}, \&
  \bibinfo{author}{{Moskalenko}, I.~V.} (\bibinfo{year}{2013}).
\newblock \bibinfo{title}{{Analytic Solution for Self-regulated Collective
  Escape of Cosmic Rays from Their Acceleration Sites}}.
\newblock {\it \bibinfo{journal}{The Astrophysical Journal}\/},  {\it
  \bibinfo{volume}{768}\/}\bibinfo{issue}{(1)}, \bibinfo{pages}{73}.
  \DOIprefix\doi{10.1088/0004-637X/768/1/73}.
  \href{http://arxiv.org/abs/1207.4728}{\tt arXiv:1207.4728}.
\bibitem[{{McKee} \& {Ostriker}(1977)}]{MO77}
\bibinfo{author}{{McKee}, C.~F.}, \& \bibinfo{author}{{Ostriker}, J.~P.}
  (\bibinfo{year}{1977}).
\newblock \bibinfo{title}{{A theory of the interstellar medium: three
  components regulated by supernova explosions in an inhomogeneous substrate.}}
\newblock {\it \bibinfo{journal}{The Astrophysical Journal}\/},  {\it
  \bibinfo{volume}{218}\/}, \bibinfo{pages}{148--169}.
  \DOIprefix\doi{10.1086/155667}.
\bibitem[{{Mestre} et~al.(2021){Mestre}, {de O{\~n}a Wilhelmi}, {Torres},
  {Holch}, {Schwanke}, {Aharonian}, {Parkinson}, {Yang} \&
  {Zanin}}]{Mestre2021}
\bibinfo{author}{{Mestre}, E.}, \bibinfo{author}{{de O{\~n}a Wilhelmi}, E.},
  \bibinfo{author}{{Torres}, D.~F.}, \bibinfo{author}{{Holch}, T.~L.},
  \bibinfo{author}{{Schwanke}, U.}, \bibinfo{author}{{Aharonian}, F.},
  \bibinfo{author}{{Parkinson}, P.~S.}, \bibinfo{author}{{Yang}, R.}, \&
  \bibinfo{author}{{Zanin}, R.} (\bibinfo{year}{2021}).
\newblock \bibinfo{title}{{Probing the hadronic nature of the gamma-ray
  emission associated with Westerlund 2}}.
\newblock {\it \bibinfo{journal}{Monthly Notices of the Royal Astronomical
  Society}\/},  {\it \bibinfo{volume}{505}\/}\bibinfo{issue}{(2)},
  \bibinfo{pages}{2731--2740}. \DOIprefix\doi{10.1093/mnras/stab1455}.
  \href{http://arxiv.org/abs/2105.09155}{\tt arXiv:2105.09155}.
\bibitem[{{Morlino} et~al.(2021){Morlino}, {Blasi}, {Peretti} \&
  {Cristofari}}]{2021MNRAS.504.6096M}
\bibinfo{author}{{Morlino}, G.}, \bibinfo{author}{{Blasi}, P.},
  \bibinfo{author}{{Peretti}, E.}, \& \bibinfo{author}{{Cristofari}, P.}
  (\bibinfo{year}{2021}).
\newblock \bibinfo{title}{{Particle acceleration in winds of star clusters}}.
\newblock {\it \bibinfo{journal}{Monthly Notices of the Royal Astronomical
  Society}\/},  {\it \bibinfo{volume}{504}\/}\bibinfo{issue}{(4)},
  \bibinfo{pages}{6096--6105}. \DOIprefix\doi{10.1093/mnras/stab690}.
  \href{http://arxiv.org/abs/2102.09217}{\tt arXiv:2102.09217}.
\bibitem[{Moskalenko et~al.(2019)Moskalenko, Johannesson \&
  Porter}]{Moskalenko:2019MU}
\bibinfo{author}{Moskalenko, I.}, \bibinfo{author}{Johannesson, G.}, \&
  \bibinfo{author}{Porter, T.} (\bibinfo{year}{2019}).
\newblock \bibinfo{title}{{GALPROP Code for Galactic Cosmic Ray Propagation and
  Associated Photon Emissions}}.
\newblock In {\it \bibinfo{booktitle}{Proceedings of 36th International Cosmic
  Ray Conference {\textemdash} PoS(ICRC2019)}\/} (p. \bibinfo{pages}{111}).
\newblock volume \bibinfo{volume}{358}.
\newblock \DOIprefix\doi{10.22323/1.358.0111}.
\bibitem[{{Muno} et~al.(2006){Muno}, {Law}, {Clark}, {Dougherty}, {de Grijs},
  {Portegies Zwart} \& {Yusef-Zadeh}}]{Muno06}
\bibinfo{author}{{Muno}, M.~P.}, \bibinfo{author}{{Law}, C.},
  \bibinfo{author}{{Clark}, J.~S.}, \bibinfo{author}{{Dougherty}, S.~M.},
  \bibinfo{author}{{de Grijs}, R.}, \bibinfo{author}{{Portegies Zwart}, S.}, \&
  \bibinfo{author}{{Yusef-Zadeh}, F.} (\bibinfo{year}{2006}).
\newblock \bibinfo{title}{{Diffuse, Nonthermal X-Ray Emission from the Galactic
  Star Cluster Westerlund 1}}.
\newblock {\it \bibinfo{journal}{The Astrophysical Journal}\/},  {\it
  \bibinfo{volume}{650}\/}\bibinfo{issue}{(1)}, \bibinfo{pages}{203--211}.
  \DOIprefix\doi{10.1086/507175}.
  \href{http://arxiv.org/abs/astro-ph/0606492}{\tt arXiv:astro-ph/0606492}.
\bibitem[{{Norman} \& {Ferrara}(1996)}]{1996ApJ...467..280N}
\bibinfo{author}{{Norman}, C.~A.}, \& \bibinfo{author}{{Ferrara}, A.}
  (\bibinfo{year}{1996}).
\newblock \bibinfo{title}{{The Turbulent Interstellar Medium: Generalizing to a
  Scale-dependent Phase Continuum}}.
\newblock {\it \bibinfo{journal}{The Astrophysical Journal}\/},  {\it
  \bibinfo{volume}{467}\/}, \bibinfo{pages}{280--306}.
  \DOIprefix\doi{10.1086/177603}.
  \href{http://arxiv.org/abs/astro-ph/9602146}{\tt arXiv:astro-ph/9602146}.
\bibitem[{{Ohm}(2016)}]{Ohm16}
\bibinfo{author}{{Ohm}, S.} (\bibinfo{year}{2016}).
\newblock \bibinfo{title}{{Starburst galaxies as seen by gamma-ray
  telescopes}}.
\newblock {\it \bibinfo{journal}{Comptes Rendus Physique}\/},  {\it
  \bibinfo{volume}{17}\/}\bibinfo{issue}{(6)}, \bibinfo{pages}{585--593}.
  \DOIprefix\doi{10.1016/j.crhy.2016.04.003}.
  \href{http://arxiv.org/abs/1601.06386}{\tt arXiv:1601.06386}.
\bibitem[{{Parizot} et~al.(2004){Parizot}, {Marcowith}, {van der Swaluw},
  {Bykov} \& {Tatischeff}}]{2004A&A...424..747P}
\bibinfo{author}{{Parizot}, E.}, \bibinfo{author}{{Marcowith}, A.},
  \bibinfo{author}{{van der Swaluw}, E.}, \bibinfo{author}{{Bykov}, A.~M.}, \&
  \bibinfo{author}{{Tatischeff}, V.} (\bibinfo{year}{2004}).
\newblock \bibinfo{title}{{Superbubbles and energetic particles in the Galaxy.
  I. Collective effects of particle acceleration}}.
\newblock {\it \bibinfo{journal}{Astronomy \& Astrophysics}\/},  {\it
  \bibinfo{volume}{424}\/}, \bibinfo{pages}{747--760}.
  \DOIprefix\doi{10.1051/0004-6361:20041269}.
  \href{http://arxiv.org/abs/astro-ph/0405531}{\tt arXiv:astro-ph/0405531}.
\bibitem[{{Ptuskin}(2011)}]{Ptuskin11}
\bibinfo{author}{{Ptuskin}, V.} (\bibinfo{year}{2011}).
\newblock \bibinfo{title}{{Cosmic ray propagation in the interstellar medium}}.
\newblock {\it \bibinfo{journal}{\memsai}\/},  {\it \bibinfo{volume}{82}\/},
  \bibinfo{pages}{858}.
\bibitem[{{Ptuskin} et~al.(2006){Ptuskin}, {Moskalenko}, {Jones}, {Strong} \&
  {Zirakashvili}}]{PMJSZ2006}
\bibinfo{author}{{Ptuskin}, V.~S.}, \bibinfo{author}{{Moskalenko}, I.~V.},
  \bibinfo{author}{{Jones}, F.~C.}, \bibinfo{author}{{Strong}, A.~W.}, \&
  \bibinfo{author}{{Zirakashvili}, V.~N.} (\bibinfo{year}{2006}).
\newblock \bibinfo{title}{{Dissipation of Magnetohydrodynamic Waves on
  Energetic Particles: Impact on Interstellar Turbulence and Cosmic-Ray
  Transport}}.
\newblock {\it \bibinfo{journal}{The Astrophysical Journal}\/},  {\it
  \bibinfo{volume}{642}\/}, \bibinfo{pages}{902--916}.
  \href{http://arxiv.org/abs/arXiv:astro-ph/0510335}{\tt
  arXiv:arXiv:astro-ph/0510335}.
\bibitem[{{Seo} et~al.(2018){Seo}, {Kang} \& {Ryu}}]{Seo2018}
\bibinfo{author}{{Seo}, J.}, \bibinfo{author}{{Kang}, H.}, \&
  \bibinfo{author}{{Ryu}, D.} (\bibinfo{year}{2018}).
\newblock \bibinfo{title}{{The Contribution of Stellar Winds to Cosmic Ray
  Production}}.
\newblock {\it \bibinfo{journal}{Journal of Korean Astronomical Society}\/},
  {\it \bibinfo{volume}{51}\/}\bibinfo{issue}{(2)}, \bibinfo{pages}{37--48}.
  \DOIprefix\doi{10.5303/JKAS.2018.51.2.37}.
  \href{http://arxiv.org/abs/1804.07486}{\tt arXiv:1804.07486}.
\bibitem[{{Strong} et~al.(2007){Strong}, {Moskalenko} \& {Ptuskin}}]{SMP2007}
\bibinfo{author}{{Strong}, A.~W.}, \bibinfo{author}{{Moskalenko}, I.~V.}, \&
  \bibinfo{author}{{Ptuskin}, V.~S.} (\bibinfo{year}{2007}).
\newblock \bibinfo{title}{{Cosmic-Ray Propagation and Interactions in the
  Galaxy}}.
\newblock {\it \bibinfo{journal}{Annual Review of Nuclear and Particle
  Science}\/},  {\it \bibinfo{volume}{57}\/}, \bibinfo{pages}{285--327}.
  \href{http://arxiv.org/abs/arXiv:astro-ph/0701517}{\tt
  arXiv:arXiv:astro-ph/0701517}.
\bibitem[{{Sun} et~al.(2020){Sun}, {Yang}, {Liang}, {Peng}, {Zhang}, {Wang} \&
  {Aharonian}}]{SunW40}
\bibinfo{author}{{Sun}, X.-N.}, \bibinfo{author}{{Yang}, R.-Z.},
  \bibinfo{author}{{Liang}, Y.-F.}, \bibinfo{author}{{Peng}, F.-K.},
  \bibinfo{author}{{Zhang}, H.-M.}, \bibinfo{author}{{Wang}, X.-Y.}, \&
  \bibinfo{author}{{Aharonian}, F.} (\bibinfo{year}{2020}).
\newblock \bibinfo{title}{{Diffuse {\ensuremath{\gamma}}-ray emission toward
  the massive star-forming region, W40}}.
\newblock {\it \bibinfo{journal}{Astronomy \& Astrophysics}\/},  {\it
  \bibinfo{volume}{639}\/}, \bibinfo{pages}{A80}.
  \DOIprefix\doi{10.1051/0004-6361/202037580}.
  \href{http://arxiv.org/abs/2006.00879}{\tt arXiv:2006.00879}.
\bibitem[{{Tatischeff} et~al.(2021){Tatischeff}, {Raymond}, {Duprat}, {Gabici}
  \& {Recchia}}]{2021arXiv210615581T}
\bibinfo{author}{{Tatischeff}, V.}, \bibinfo{author}{{Raymond}, J.~C.},
  \bibinfo{author}{{Duprat}, J.}, \bibinfo{author}{{Gabici}, S.}, \&
  \bibinfo{author}{{Recchia}, S.} (\bibinfo{year}{2021}).
\newblock \bibinfo{title}{{The origin of Galactic cosmic rays as revealed by
  their composition}}.
\newblock {\it \bibinfo{journal}{Monthly Notices of the Royal Astronomical
  Society}\/},  {\it \bibinfo{volume}{508}\/}\bibinfo{issue}{(1)},
  \bibinfo{pages}{1321--1345}. \DOIprefix\doi{10.1093/mnras/stab2533}.
  \href{http://arxiv.org/abs/2106.15581}{\tt arXiv:2106.15581}.
\bibitem[{{Tolksdorf} et~al.(2019){Tolksdorf}, {Grenier}, {Joubaud} \&
  {Schlickeiser}}]{2019ApJ...879...66T}
\bibinfo{author}{{Tolksdorf}, T.}, \bibinfo{author}{{Grenier}, I.~A.},
  \bibinfo{author}{{Joubaud}, T.}, \& \bibinfo{author}{{Schlickeiser}, R.}
  (\bibinfo{year}{2019}).
\newblock \bibinfo{title}{{Cosmic Rays in Superbubbles}}.
\newblock {\it \bibinfo{journal}{The Astrophysical Journal}\/},  {\it
  \bibinfo{volume}{879}\/}\bibinfo{issue}{(2)}, \bibinfo{pages}{66}.
  \DOIprefix\doi{10.3847/1538-4357/ab24c6}.
\bibitem[{{Toptygin}(1985)}]{1985crim.book.....T}
\bibinfo{author}{{Toptygin}, I.~N.} (\bibinfo{year}{1985}).
\newblock {\it \bibinfo{title}{{Cosmic rays in interplanetary magnetic
  fields}}\/}.
\newblock \bibinfo{publisher}{Springer, Dordrecht}.
\bibitem[{{Vargas {\'A}lvarez} et~al.(2013){Vargas {\'A}lvarez}, {Kobulnicky},
  {Bradley}, {Kannappan}, {Norris}, {Cool} \& {Miller}}]{Alvarez2013}
\bibinfo{author}{{Vargas {\'A}lvarez}, C.~A.}, \bibinfo{author}{{Kobulnicky},
  H.~A.}, \bibinfo{author}{{Bradley}, D.~R.}, \bibinfo{author}{{Kannappan},
  S.~J.}, \bibinfo{author}{{Norris}, M.~A.}, \bibinfo{author}{{Cool}, R.~J.},
  \& \bibinfo{author}{{Miller}, B.~P.} (\bibinfo{year}{2013}).
\newblock \bibinfo{title}{{The Distance to the Massive Galactic Cluster
  Westerlund 2 from a Spectroscopic and HST Photometric Study}}.
\newblock {\it \bibinfo{journal}{The Astronomical Journal}\/},  {\it
  \bibinfo{volume}{145}\/}\bibinfo{issue}{(5)}, \bibinfo{pages}{125}.
  \DOIprefix\doi{10.1088/0004-6256/145/5/125}.
  \href{http://arxiv.org/abs/1302.0863}{\tt arXiv:1302.0863}.
\bibitem[{{Vieu} et~al.(2020){Vieu}, {Gabici} \&
  {Tatischeff}}]{2020MNRAS.494.3166V}
\bibinfo{author}{{Vieu}, T.}, \bibinfo{author}{{Gabici}, S.}, \&
  \bibinfo{author}{{Tatischeff}, V.} (\bibinfo{year}{2020}).
\newblock \bibinfo{title}{{Particle acceleration at colliding shock waves}}.
\newblock {\it \bibinfo{journal}{Monthly Notices of the Royal Astronomical
  Society}\/},  {\it \bibinfo{volume}{494}\/}\bibinfo{issue}{(3)},
  \bibinfo{pages}{3166--3176}. \DOIprefix\doi{10.1093/mnras/staa799}.
  \href{http://arxiv.org/abs/2003.03411}{\tt arXiv:2003.03411}.
\bibitem[{{Wright} et~al.(2015){Wright}, {Drew} \& {Mohr-Smith}}]{Wright2015}
\bibinfo{author}{{Wright}, N.~J.}, \bibinfo{author}{{Drew}, J.~E.}, \&
  \bibinfo{author}{{Mohr-Smith}, M.} (\bibinfo{year}{2015}).
\newblock \bibinfo{title}{{The massive star population of Cygnus OB2}}.
\newblock {\it \bibinfo{journal}{Monthly Notices of the Royal Astronomical
  Society}\/},  {\it \bibinfo{volume}{449}\/}\bibinfo{issue}{(1)},
  \bibinfo{pages}{741--760}. \DOIprefix\doi{10.1093/mnras/stv323}.
  \href{http://arxiv.org/abs/1502.05718}{\tt arXiv:1502.05718}.
\bibitem[{{Yang} et~al.(2018){Yang}, {de O{\~n}a Wilhelmi} \&
  {Aharonian}}]{YangWd2}
\bibinfo{author}{{Yang}, R.-z.}, \bibinfo{author}{{de O{\~n}a Wilhelmi}, E.},
  \& \bibinfo{author}{{Aharonian}, F.} (\bibinfo{year}{2018}).
\newblock \bibinfo{title}{{Diffuse {\ensuremath{\gamma}}-ray emission in the
  vicinity of young star cluster Westerlund 2}}.
\newblock {\it \bibinfo{journal}{Astronomy \& Astrophysics}\/},  {\it
  \bibinfo{volume}{611}\/}, \bibinfo{pages}{A77}.
  \DOIprefix\doi{10.1051/0004-6361/201732045}.
  \href{http://arxiv.org/abs/1710.02803}{\tt arXiv:1710.02803}.

\end{thebibliography}

\end{document}